\documentclass[aps,prd,onecolumn,showpacs,showkeys,amsmath,amssymb]{revtex4}
\usepackage{graphicx}
\usepackage{dcolumn}
\usepackage{bm}
\usepackage{CJK}
\usepackage{CJKfntef}
\def\half{\frac{1}{2}}

\def\inner#1#2{{\bm #1}\cdot {\bm #2}}

\def\half{\frac{1}{2}}

\begin{document}
\title{Baryon Fields with $U_L(3)\times U_R(3)$ Chiral Symmetry:
Axial Currents of Nucleons and Hyperons}
%
\author{Hua-Xing Chen$^{1}$}
\email{hxchen@rcnp.osaka-u.ac.jp}
\author{V. Dmitra\v sinovi\' c$^2$}
\email{dmitra@vinca.rs}
\author{Atsushi Hosaka$^{3}$}
\email{hosaka@rcnp.osaka-u.ac.jp}
\affiliation{$^1$Department of Physics and State Key Laboratory of
Nuclear Physics and Technology Peking University, Beijing 100871,
China \\ $^2$ Vin\v ca Institute
of Nuclear Sciences, lab 010, P.O.Box 522, 11001 Beograd, Serbia \\
$^3$ Research Center for Nuclear Physics, Osaka University,
Ibaraki 567--0047, Japan}

\begin{abstract}
We use the conventional $F$ and $D$ octet and decimet generator
matrices to reformulate chiral properties of local (non-derivative)
and one-derivative non-local fields of baryons consisting of three
quarks with flavor $SU(3)$ symmetry that were expressed in $SU(3)$
tensor form in Ref.~\cite{Chen:2008qv}. We show explicitly the
chiral transformations of the $[(6,3)\oplus(3,6)]$ chiral multiplet
in the ``$SU(3)$ particle basis'', for the first time to our
knowledge, as well as those of the $(3,\overline{3}) \oplus
(\overline{3}, 3)$, $(8,1) \oplus (1, 8)$ multiplets, which have
been recorded before in Refs.~\cite{Lee:1968,Bardeen:1969ra}. We
derive the vector and axial-vector Noether currents, and show
explicitly that their zeroth (charge-like) components close the
$SU_L(3) \times SU_R(3)$ chiral algebra. We use these results to
study the effects of mixing of (three-quark) chiral multiplets on
the axial current matrix elements of hyperons and nucleons. We show,
in particular, that there is a strong correlation, indeed a definite
relation between the flavor-singlet (i.e. the zeroth), the isovector
(the third) and the eighth flavor component of the axial current,
which is in decent agreement with the measured ones.
\end{abstract}
\maketitle
\pagenumbering{arabic}
%

\section{Introduction}
\label{Intro}

Axial current ``coupling constants'' of the baryon flavor
octet~\cite{Okun:1982ap} are well known by now, see
Ref.~\cite{Yamanishi:2007zza}\footnote{for history and other
references, see Chapter 6.7 of Okun's book~\cite{Okun:1982ap} and
PDG tables~\cite{Amsler:2008zzb}}. The zeroth (time-like) components
of these axial currents are generators of the $SU_L(3) \times
SU_R(3)$ chiral symmetry that is one of the fundamental symmetries
of QCD. The general flavor $SU_F(3)$ symmetric form of the nucleon
axial current contains two free parameters, the so called $F$ and
$D$ couplings, which are empirically determined as $F$=$0.459 \pm
0.008$ and $D$=$0.798 \pm 0.008$, see Ref.~\cite{Yamanishi:2007zza}.
The conventional models of (linearly realized) chiral $SU_L(3)
\times SU_R(3)$ symmetry, Refs.~\cite{Lee:1968,Bardeen:1969ra}, on
the other hand appear to fix these parameters at either
($F$=0,$D$=1), which case goes by the name of $[(3,\overline{3})
\oplus (\overline{3}, 3)]$, or at ($F$=1,$D$=0), which case goes by
the name of $[(8,1)\oplus (1,8)]$ representation. Both of these
chiral representations suffer from the shortcoming that
$F$+$D$=1$\neq g_A^{(3)}=$1.267 without derivative couplings. But,
even with derivative interactions, one cannot change the value of
the vanishing coupling, e.g. of $F$=0, in $[(3,\overline{3}) \oplus
(\overline{3}, 3)]$, or of $D$=0, in $[(8,1)\oplus (1,8)]$. Rather,
one can only renormalize the non-vanishing coupling to 1.267.

Attempts at a reconciliation of the measured values of axial
couplings with the (broken) $SU_L(3) \times SU_R(3)$ chiral
symmetry go back at least 40
years~\cite{Hara:1965,Lee:1968,Bardeen:1969ra,Harari:1966yq,Harari:1966jz,Gerstein:1966zz,Weinberg:1969hw},
but, none have been successful to our knowledge thus far. As noted
above, perhaps the most troublesome problem are the $SU(3)$ axial
current's $F$,$D$ values, which problem has repercussions for the
meson-baryon interaction $F$,$D$ values, with far-reaching
consequences for hyper-nuclear physics and even astrophysics.
Another, perhaps equally important and difficult problem is that
of the flavor-singlet axial coupling of the nucleon
\cite{Bass:2007zzb}. This is widely thought of as being
disconnected from the $F$,$D$ problem, but we shall show that the
three-quark interpolating fields cast some perhaps unexpected
light on this problem. We shall attack both of these problems from
Weinberg's~\cite{Weinberg:1969hw} point of view, {\it viz.} chiral
representation mixing, extended to the $SU_L(3) \times SU_R(3)$
and $U_L(1) \times U_R(1)$ chiral symmetries, with added input
from three-quark baryon interpolating fields~\cite{Chen:2008qv}
that are ordinarily used in QCD calculations.

The basic idea is simple: a mixture of two baryon fields belonging
to different chiral representations/multiplets has axial couplings
that lie between the extreme values determined by the two chiral
multiplets that are being mixed, and depend on the mixing angle, of
course. Weinberg used this idea to fit the iso-vector axial coupling
of the nucleon using the $[(1/2,0)\oplus (0,1/2)]$ and
$[(1,1/2)\oplus (1/2,1)]$ multiplets of the $SU_L(2) \times SU_R(2)$
chiral symmetry, but the same idea may be used on any baryon
belonging to the same octet, e.g. for the $\Lambda, \Sigma$ and
$\Xi$ hyperons. In other words, the $F$ and $D$ values of the
mixture can be determined from the $F$ and $D$ values of the
$SU_L(3) \times SU_R(3)$ representations corresponding to the
$[(1/2,0)\oplus (0,1/2)]$ and $[(1,1/2)\oplus (1/2,1)]$ multiplets,
{\it viz.} $[(3,\overline{3}) \oplus (\overline{3}, 3)]$ or
$[(8,1)\oplus (1,8)]$, and $[(6,3)\oplus (3,6)]$, respectively. The
same principle holds for the $U_L(1) \times U_R(1)$ symmetry
``multiplets'' and the value(s) of the flavor singlet axial charge.

The $SU_L(3) \times SU_R(3)$ and $U_L(1) \times U_R(1)$ chiral
transformation properties of three-quark baryon interpolating
fields, that are commonly used in various QCD (lattice, sum rules)
calculations, and that have recently been determined in
Ref.~\cite{Chen:2008qv} will be used here as input into the chiral
mixing formalism, so as to deduce as much phenomenological
information about the axial currents of hyperons and nucleons as
possible. As a result we find three ``optimal'' scenarios all with
identical $F$,$D$ values (see Sect. \ref{sect:mix}).

First we recast our previous results~\cite{Chen:2008qv} into the
language that is conventional for axial currents, i.e. in terms of
octet $F$ and $D$ couplings. A large part of the present paper is
devoted to this notational conversion (change of basis) and the
subsequent check whether and how the resulting chiral charges
actually satisfy the $SU(3)\times SU(3)$ chiral algebra. That is a
non-trivial task for the $[(3,\overline{3}) \oplus (\overline{3},
3)]$ and $(6,3)\oplus(3,6)$ representations, because they involve
off-diagonal terms, and in the latter case one of the diagonal terms
in the axial current is multiplied by a fractional coefficient, that
appears to spoil the closure of the $SU(3)\times SU(3)$ chiral
algebra; the off-diagonal terms in the axial current make crucial
contributions that restore the closure. Thus, the afore-mentioned
fractional coefficient is uniquely determined.

We use these results to study the effects of mixing of
(three-quark) chiral multiplets on the axial current matrix
elements of hyperons and nucleons. We show, in particular, that
there is a strong correlation between the flavor-singlet (i.e. the
zeroth), the isovector (the third) and the eighth flavor component
of the axial current. There are, in principle, three independent
observables here: the flavor-singlet (i.e. the zeroth), the
isovector (the third) and the eighth flavor component of the axial
current of the nucleon. By fitting just one mixing angle to one of
these values, e.g. the (best known) isovector coupling, we predict
the other two. These predictions may differ widely depending on
the field that one assumes to be mixed with the $(6,3)\oplus(3,6)$
field (which must be present if the isovector axial coupling has
any chance of being fit). If one assumes mixing of three fields
(again, always keeping the $(6,3)\oplus(3,6)$ as one of the three)
and fits the flavor-singlet and the isovector axial couplings,
then one finds a {\it unique} prediction for the $F$, $D$ values,
which is in decent agreement with the measured ones, modulo
$SU(3)$ symmetry breaking corrections, which may be important (for
a recent fit, see Ref.~\cite{Yamanishi:2007zza}). The uniqueness
of this result is a consequence of a remarkable relation,
$g_{A}^{(0)}=3F-D$ that holds for all three (five) chiral
multiplets involved here, and which leads to the relation:
$g_{A}^{(0)} = \sqrt{3} g_{A}^{(8)}$, see Sect. \ref{sect:mix}.

Most of the ideas used in this paper, such as that of chiral
multiplet mixing, have been presented in mid- to late 1960's,
Refs.
\cite{Harari:1966yq,Weinberg:1969hw,Harari:1966jz,Gerstein:1966zz},
with the (obvious) exception of the use of QCD interpolating
fields, which arrived only a decade afterwards/later, and the
(perhaps less obvious) question of baryons' flavor-singlet axial
current (a.k.a. the $U_A(1)$), which was (seriously) raised yet
another decade later.

The present paper consists of five parts: after the present
Introduction, in Sect.~\ref{sect:su3} we define the $SU(3)\times
SU(3)$ chiral transformations of three-quark baryon fields, with
special emphasis on the $SU(3)$ phase conventions 
that ensure standard $SU(2)$ isospin conventions for the isospin
sub-multiplets, and we define the ($SU(3)$ symmetric) vector and
axial-vector Noether currents of three-quark baryon fields. In
Sect. \ref{sect:SU(3)chiral algebra} we prove the closure of the
chiral $SU_L(3) \times SU_R(3)$ algebra. In Sect.~\ref{sect:mix}
we apply chiral mixing formalism to the hyperons' axial currents
and discuss the results. Finally, in Sect.~\ref{sect:summary} we
offer a summary and an outlook on future developments.



\section{$SU(3)\times SU(3)$ Chiral Transformations of Three-quark
Baryon Fields and their Noether Currents} \label{sect:su3}

We must make sure that our conventions ensure that identical
isospin multiplets in different $SU(3)$ multiplets, such as the
octet and the decuplet, have identical isospin
algebras/generators. That is a relatively simple matter of
definition, but was not the case with the octet conventions used
in Ref.~\cite{Chen:2008qv}. Our new definitions of the octet and
decuplet fields avoid these problems.

\subsection{Octet and Decuplet State Definition}
\label{sect:Octet_Decuplet_Definition}

The new $\Xi^-$ wave function comes with a minus sign: that is
precisely the convention used in Eqs. (18) and (19) in Sect. 18 of
Gasiorowicz's textbook~\cite{gas}. But then we must also adjust
the $8\times10$ $SU(3)$-spurion matrices for this modification.
\begin{eqnarray}
& & \Sigma^{\mp} \sim {\pm 1 \over \sqrt{2}}(N^1 \pm i N^2 ) \, ,
\; \; \; N^3 \sim \Sigma^0 \, , \; \; \; N^8 \sim \Lambda_8 \, , \\
\nonumber & & \left(\begin{array}{c} ~\Xi^- \\ p
\end{array}\right) \sim {\mp 1 \over \sqrt{2}}(N^4 \pm i N^5) ,\;
\; \left(\begin{array}{c} ~\Xi^0 \\ n
\end{array}\right) \; \sim \, {1 \over \sqrt{2}}(N^6 \pm i N^7) .
\label{def:octet_state}
\end{eqnarray}
\begin{eqnarray}
\left (\begin{array}{c} p \\ n \\ \Sigma^+
\\ \Sigma^0 \\ \Sigma^- \\ \Xi^0 \\ \Xi^- \\ \Lambda_8
\end{array}
\right ) &=& \left(
\begin{array}{llllllll}
 0 & 0 & 0 & \frac{1}{\sqrt{2}} & \frac{-i}{\sqrt{2}} & 0 & 0 & 0 \\
 0 & 0 & 0 & 0 & 0 & \frac{1}{\sqrt{2}} & \frac{-i}{\sqrt{2}} & 0 \\
 \frac{-1}{\sqrt{2}} & \frac{i}{\sqrt{2}} & 0 & 0 & 0 & 0 & 0 & 0 \\
 0 & 0 & 1 & 0 & 0 & 0 & 0 & 0 \\
 \frac{1}{\sqrt{2}} & \frac{i}{\sqrt{2}} & 0 & 0 & 0 & 0 & 0 & 0 \\
 0 & 0 & 0 & 0 & 0 & \frac{1}{\sqrt{2}} & \frac{i}{\sqrt{2}} & 0 \\
 0 & 0 & 0 & \frac{-1}{\sqrt{2}} & \frac{-i}{\sqrt{2}} & 0 & 0 & 0 \\
 0 & 0 & 0 & 0 & 0 & 0 & 0 & 1
\end{array}
\right) \left (\begin{array}{c} N^1 \\ N^2 \\ N^3
\\ N^4 \\ N^5 \\ N^6 \\ N^7 \\ N^8
\end{array}
\right ) \, ,
\end{eqnarray}
or put them into the $3 \times 3$ baryon matrix as follows
\begin{eqnarray}
{\mathfrak N} = \left ( \begin{array} {c c c} {\Sigma^0 \over
\sqrt{2}} + { \Lambda^8 \over \sqrt{6}} & - \Sigma^+ & p
\\ \Sigma^- & - {\Sigma^0 \over \sqrt{2}} + { \Lambda^8
\over \sqrt{6}} & n
\\ - \Xi^- & \Xi^0 & - {2\over\sqrt{6}} \Lambda^8
\end{array} \right )
\label{eq:B} \, .
\end{eqnarray}
Note the minus signs in front of $\Xi^-$ and $\Sigma^+$. We also use
a new normalization of the decuplet fields:
\begin{eqnarray}
&& { \Delta^1 } \sim  - {1 \over \sqrt3}\Delta^{++} \, , { \Delta^7
}
\sim - {1 \over \sqrt3} \Delta^{-} \, , { \Delta^{10} } \sim
- {1 \over \sqrt3} \Omega^{-} \, , \\
\nonumber &&  \Delta^2  \sim  - \Delta^{+} \, ,  \Delta^4 \sim -
\Delta^{0} \, ,  \Delta^3  \sim - \Sigma^{*+} \, ,  \Delta^8 \sim -
\Sigma^{*-} \, , \Delta^6 \sim
- \Xi^{*0} \, , \Delta^9 \sim - \Xi^{*-} \, , \\
\nonumber && \Delta^5 \sim - \sqrt{2} \Sigma^{*0}
\label{def:decimet_state new}
\end{eqnarray}
For the singlet $\Lambda$, we use the normalization:
\begin{equation}
\Lambda_{1} = \Lambda_{phy} = { 2 \sqrt 2 \over \sqrt 3 } \Lambda \,
.
\end{equation}
For simplicity, we will just use $\Lambda_1$ instead of
$\Lambda_{phy}$ in the following sections.

We define the flavor octet and decuplet matrices/column vectors as
\begin{eqnarray}
N &=& (p, n, \Sigma^+, \Sigma^0, \Sigma^-, \Xi^0, \Xi^-,
\Lambda_8)^T \, , \\
\Delta &=& (\Delta^{++}, \Delta^+, \Delta^0,
\Delta^- , \Sigma^{*+}, \Sigma^{*0}, \Sigma^{*-} , \Xi^{*0},
\Xi^{*-}, \Omega)^T
\end{eqnarray}
In our previous paper, Ref.~\cite{Chen:2008qv}, we found that the
baryon interpolating fields $N_+^a = N^a_1 + N^a_2$ belong to the
chiral representation $(\mathbf{8}, \mathbf{1}) \oplus (\mathbf{1},
\mathbf{8})$; $\Lambda$ and $N_-^a = N^a_1 - N^a_2$ belong to the
chiral representation $(\mathbf{3}, \mathbf{\overline{3}}) \oplus
(\mathbf{\overline{3}}, \mathbf{3})$; $N^{a}_\mu$ and
$\Delta^P_{\mu}$ belong to the chiral representation $(\mathbf{6},
\mathbf{3}) \oplus (\mathbf{3}, \mathbf{6})$; and
$\Delta^P_{\mu\nu}$ belong to the chiral representation
$(\mathbf{10}, \mathbf{1}) \oplus (\mathbf{1}, \mathbf{10})$. Here
$N^a_1$ and $N^a_2$ are the two independent kinds of nucleon fields.
$N^a_1$ contains the ``scalar diquark'' and $N^a_2$ contains the
``pseudoscalar diquark''. Moreover, we calculated their chiral
transformations in Ref.~\cite{Chen:2008qv}. That form, however, is
not conventionally used for the axial currents. So in the following
subsections, we use different conventions, listed above, and display
the chiral transformations in these bases.

\subsection{Chiral Transformations of Three-Quark Interpolating Fields}
\label{chi_tra_three}

\subsubsection{$(8, 1) \oplus (1, 8)$ Chiral Transformations}

This chiral representation contains the flavor octet representation
$\mathbf{8}$. For the octet baryon field $N^a$ ($a=1,\cdots,8$),
chiral transformations are given by:
\begin{eqnarray}
\delta_5^{\vec b} N_+ &=& i \gamma_5 b^a {\bf F}_{(8)}^{a} N_+
\label{e:N81} \, , \
\end{eqnarray}
The $SU(3)$-spurion matrices ${\bf F}_{(8)}^{a}$ are listed in the
Appendix~\ref{s:Veljko8x10}. This corresponds to the chiral
transformations of Ref.~\cite{Chen:2008qv}:
\begin{eqnarray}
\delta_5^{\vec{b}} (N^a_1 + N^a_2) &=& \gamma_5 b^b f^{bac} ( N^c_1
+ N^c_2 ) \, . \nonumber
\end{eqnarray}
The coefficients $f^{abc}$ are the standard antisymmetric
``structure constants'' of $SU(3)$. For completeness' sake, we show
the following equation which defines the $f$ and $d$ coefficients
\begin{eqnarray}
\lambda^a_{AB} \lambda^b_{BC} &=& (\lambda^a \lambda^b)_{AC} =
{1\over2} \{ \lambda^a, \lambda^b \}_{AC} + {1\over2} [ \lambda^a,
\lambda^b ]_{AC} \nonumber \\
&=& {2\over3} \delta^{ab} \delta_{AC} + (d^{abc} + i f^{abc})
\lambda^c_{AC} \, .
\end{eqnarray}

\subsubsection{$(3, \overline{3}) \oplus (\overline{3}, 3)$ Chiral
Transformations}

This chiral representation contains the flavor octet and singlet
representations $\mathbf{\bar 3} \otimes \mathbf{3} = \mathbf{8}
\oplus \mathbf{1}$ $\sim(N^a, \Lambda)$. These two flavor
representations are mixed under chiral transformations as
\begin{eqnarray}
\delta_5^{\vec b} \Lambda_1 &=& i \gamma_5 b^a \sqrt{2 \over 3}
{\rm \bf  T}^a_{1/8} N_- \nonumber \\
\delta_5^{\vec b} N_- &=& i \gamma_5 b^a \left({\rm {\bf D}}^{a}
N_- + \sqrt{2 \over 3} {\rm \bf  T}^{a\dagger}_{1/8} \Lambda_1
\right)
 \label{e:N33} \, . \
\end{eqnarray}
where ${\rm {\bf D}}^{a}$ are defined in the
Appendix~\ref{sect:Octet Generators Veljko}. The $SU(3)$-spurion
matrices ${\rm \bf T}^{a}_{1/8}$ have the following properties
\begin{eqnarray}
{\rm \bf T}^{a}_{1/8} {\rm \bf T}^{a\dagger}_{1/8} &=& 8
\nonumber \\
{\rm \bf T}^{a\dagger}_{1/8} {\rm \bf T}^{a}_{1/8} &=& {\mathbf
1}_{8\times8} \, , \label{e:app32b}
\end{eqnarray}
and are listed in the Appendix~\ref{s:Veljko8x1}. Here ${\mathbf
1}_{8\times8}$ is a unit matrix of $8\times8$ dimensions.

\subsubsection{$(6, 3)\oplus(3, 6)$ Chiral Transformations}

This chiral representation contains flavor octet and decuplet
representations $\mathbf{6} \otimes \mathbf{3} = \mathbf{8} \oplus
\mathbf{10}$ $\sim(N^a, \Delta^b)$. For their chiral transformations
we use the results from Ref.~\cite{Chen:2008qv}, where they were
expressed in terms of coefficients $g$, $g^\prime$,
$g^{\prime\prime}$ and $g^{\prime\prime\prime}$ that were tabulated
in Table~II. For off-diagonal terms (between octet and decuplet),
there is a (new) factor $1\over6$, which comes from the different
normalization of octet and decuplet. Here we show the final result:
\begin{eqnarray}
\delta_5^{\vec b} N &=& i \gamma_5 b^a \left({\rm ({\bf D}^{a} +
{2\over3} {\bf F}_{(8)}^{a})} N + \frac{2}{\sqrt{3}} {\rm {\bf
T}}^{a} \Delta \right) \label{e:N36} \, , \
\nonumber \\
\delta_5^{\vec b} \Delta &=& i \gamma_5 b^a \left(\frac{2}{\sqrt{3}}
{\rm {\bf T}}^{a \dagger} N + \frac{1}{3} {\rm {\bf F}}_{(10)}^{a}
\Delta \right) \label{e:delta36} \, . \
\end{eqnarray}
These $SU(3)$-spurion matrices ${\bf T}^{a}$ (sometimes we use
${\bf T}^{a}_{10/8}$) and ${\bf F}_{(10)}^{a}$ have the following
properties
\begin{eqnarray}
{\rm {\bf F}}_{(10)}^{a} &=& - \, i \, f^{abc}{\bf T}^{b \dagger}
{\bf T}^{c} \,
\nonumber \\
{\bf T}^{a} {\bf T}^{a \dagger} &=&  \, \frac52 \ {\mathbf
1}_{8\times8}
\nonumber \\
{\bf T}^{a \dagger} {\bf T}^{a} &=&  \, 2 \ {\mathbf
1}_{10\times10} \, , \label{e:app32c}
\end{eqnarray}
These transition matrices ${\bf T}^{c}$ and the decuplet
generators ${\rm {\bf F}}_{(10)}^{a}$ are listed in
Appendices~\ref{s:Veljko8x10} and \ref{sect:Decimet Generators},
respectively.

\subsection{Noether Currents of the Chiral $SU_L(3) \times SU_R(3)$
Symmetry} \label{ssect:SU(3)Neother currents}

The chiral $SU_L(3) \times SU_R(3)$ transformations of the baryon
fields $B_{i}$ define eight components of the baryon isovector axial
current ${\bf J}_{\mu 5}^a$, by way of Noether's theorem:
\begin{eqnarray}
- \inner{b}{J_{\mu 5}} &=& \sum_{i} \frac{\partial {\cal
L}}{\partial \partial^{\mu} B_{i}} \delta_{5}^{\vec{b}} B_{i}.
\label{e:Noether1} \
\end{eqnarray}
Similarly, the flavor $SU(3)$ transformations $\delta^{\vec{a}}
B_{i}$ define the Lorentz-vector Noether (flavor) current
\begin{eqnarray}
- \inner{a}{J_{\mu}} &=& \sum_{i} \frac{\partial {\cal
L}}{\partial
\partial^{\mu} B_{i}} \delta^{\vec{a}} B_{i}. \label{e:Noether
vec} \
\end{eqnarray}

\subsubsection{The Axial Current in the $(8,1) \oplus (1, 8)$
Multiplet} \label{ssect:axcurr81a}

Eqs.~(\ref{e:N81}), the chiral $SU_L(3) \times SU_R(3)$
transformation rules of the $B_{i}= N_{+}^{i}$ baryons in the $(8,1)
\oplus (1, 8)$ chiral multiplet, define the eight components of the
(hyperon) flavor octet axial current ${\bf J}_{\mu 5}^a$, by way of
Noether's theorem, Eq.~(\ref{e:Noether1}), where $B_{i}$ are the
octet $N^{i}$ baryon fields. The axial current ${\bf J}_{\mu 5}$ is
\begin{eqnarray}
{\bf J}_{\mu 5}^{a} &=& \,\overline{N} \gamma_{\mu} \gamma_{5}
{\bf F}_{(8)}^{a} N ~. \label{e:axi1SU(3)81} \
\end{eqnarray}
Here ${\bf F}_{(8)}^i$ are the $SU(3)$ octet matrices/generators.
The Lorentz vector Noether (flavor-octet) current in this multiplet
reads
\begin{eqnarray}
{\bf J}_{\mu}^{a} &=& \overline{N} \gamma_{\mu} \, {\bf
F}_{(8)}^{a} \,N  ~, \label{e:vec1SU(3)81} \
\end{eqnarray}
which are valid if the interactions do not contain derivatives.

\subsubsection{The Axial Current in the
$(\overline{3}, 3) \oplus (3,\overline{3})$ Multiplet}
\label{ssect:axcurr33a}

Eqs.~(\ref{e:N33}), the chiral $SU_L(3) \times SU_R(3)$
transformation rules of the $B_{i}=(N_{-}^{i},\,\Lambda)$ baryons in
the $(\overline{3}, 3) \oplus (3,\overline{3})$ chiral multiplet,
define the eight components of the (hyperon) flavor octet axial
current ${\bf J}_{\mu 5}^a$, by way of Noether's theorem,
Eq.~(\ref{e:Noether1}), where $B_{i}$ are the flavor octet
$N_{-}^{i}$ and the flavor singlet $\Lambda_{1}$ baryon fields. The
axial current ${\bf J}_{\mu 5}$ is
\begin{eqnarray}
{\bf J}_{\mu 5}^{a} &=& \,\overline{N} \gamma_{\mu} \gamma_{5}
\left({\rm {\bf D}}^{a} N + \sqrt{2 \over 3} {\rm \bf
T}^{a\dagger}_{1/8} \Lambda_1 \right)
\nonumber \\
&+& \, \overline{\Lambda}_1 \gamma_{\mu} \gamma_{5} \sqrt{2 \over
3} {\rm \bf  T}^a_{1/8} N ~. \label{e:axi1SU(3)33} \
\end{eqnarray}
Here ${\bf D}^i$ are the $SU(3)$ octet matrices/generators. The
Lorentz vector Noether (flavor-octet) current in this multiplet
reads
\begin{eqnarray}
{\bf J}_{\mu}^{a} &=& \overline{N} \gamma_{\mu} \, {\bf
F}_{(8)}^{a} \,N ~. \label{e:vec1SU(3)33} \
\end{eqnarray}

\subsubsection{Axial Current in the $(3,6) \oplus (6, 3)$ Multiplet}
\label{ssect:axcurr36a}

The chiral $SU_L(3) \times SU_R(3)$ transformation rules of the
$B_{i}=(N^{i},\,\Delta^{j})$ baryons, Eqs.~(\ref{e:N36}), in the
$(3,6) \oplus (6, 3)$ chiral multiplet, define the eight
components of the (hyperon) flavor octet axial current ${\bf
J}_{\mu 5}^a$, by way of Noether's theorem
(Eq.~(\ref{e:Noether1})), where $B_{i}$ are the octet $N^{i}$ and
the decuplet $\Delta^{j}$ baryon fields. The axial current ${\bf
J}_{\mu 5}$ is
\begin{eqnarray}
{\bf J}_{\mu 5}^{a} &=& \,\overline{N} \gamma_{\mu} \gamma_{5}
\left({\rm ({\bf D}^{a} + {2\over3} {\bf F}_{(8)}^{a})} N +
\frac{2}{\sqrt{3}} {\rm {\bf T}}^{a} \Delta \right)
\nonumber \\
&+& \, \overline{\Delta} \gamma_{\mu} \gamma_{5}
\left(\frac{2}{\sqrt{3}} {\rm {\bf T}}^{a \dagger} N + \frac{1}{3}
{\rm {\bf F}}_{(10)}^{a} \Delta \right) ~. \label{e:axi1SU(3)63} \
\end{eqnarray}
Here ${\bf D}^i$ and ${\bf F}_{(8)}^i$ are the $SU(3)$ octet
matrices/generators ${\bf D}^a$ and ${\bf F}_{(8)}^a$, respectively,
${\bf F}_{(10)}^i$ are the $SU(3)$ decuplet generators, and ${\bf
T}^i$ are the so-called $SU(3)$-spurion matrices. The Lorentz vector
Noether (flavor-octet) current in this multiplet reads
\begin{eqnarray}
{\bf J}_{\mu}^{a} &=& \left(\overline{N} \gamma_{\mu} \, {\bf
F}_{(8)}^{a} \,N \right) + \left(\overline{\Delta} \gamma_{\mu} \,
{\bf F}_{(10)}^{a} \,\Delta \right) ~. \label{e:vec1SU(3)63} \
\end{eqnarray}

\section{Closure of the chiral $SU_L(3) \times SU_R(3)$ algebra}
\label{sect:SU(3)chiral algebra}

The $SU(3)$ vector charges $Q^a = \int d {\bf x}J_{0}^a (t,{\bf x})$
defined by Eq.~(\ref{e:Noether vec}), together with the axial
charges $Q_{5}^a = \int d {\bf x}J_{05}^a (t,{\bf x})$ defined by
Eq.~(\ref{e:Noether1}) ought to close the chiral algebra
\begin{eqnarray}
\left[Q^a , Q^b \right] &=& i f^{abc} Q^c
\label{e:cc3VVV} \\
\left[Q_{5}^a , Q^b \right] &=& i f^{abc} Q_{5}^c
\label{e:cc3AVA} \\
\left[Q_{5}^a , Q_{5}^b \right] &=& i f^{abc} Q^c~.
\label{e:cc3AAA} \
\end{eqnarray}
where $f^{abc}$ are the SU(3) structure constants.
Eqs.~(\ref{e:cc3VVV}) and (\ref{e:cc3AVA}) usually hold
automatically, as a consequence of the canonical (anti)commutation
relations between Dirac baryon fields $B_{i}$, whereas
Eq.~(\ref{e:cc3AAA}) is not trivial for the chiral multiplets that
are different from the $[(8,1) \oplus (1,8)]$, because of the
(nominally) fractional axial charges and the presence of the
off-diagonal components. When taking a matrix element of
Eq.~(\ref{e:cc3AAA}) by baryon states in a certain chiral
representation, the axial charge mixes different flavor states
within the same chiral representation. This is an algebraic version
of the Adler-Weisburger sum rule~\cite{Weinberg:1969hw}. In the
following we shall check and confirm the validity of
Eq.~(\ref{e:cc3AAA}) in the three multiplets of SU(3)$_L
\times$SU(3)$_R$.

\subsection{Closure of the Chiral $SU_L(3) \times SU_R(3)$ Algebra
in the $(8,1) \oplus (1, 8)$ Multiplet} \label{ssect:closure81a}

Due to the absence of fractional coefficients in the $(8,1) \oplus
(1, 8)$ multiplet's axial charge $Q_{5}^a = \int d {\bf x}J_{05}^a
(t,{\bf x})$ defined by the current given in
Eq.~(\ref{e:axi1SU(3)81}), the vector charge $Q^a = \int d {\bf
x}J_{0}^a (t,{\bf x})$ defined by the current given in
Eq.~(\ref{e:vec1SU(3)81}) and the axial charge close the chiral
algebra defined by Eqs.~(\ref{e:cc3VVV}), (\ref{e:cc3AVA}) and
(\ref{e:cc3AAA}). The same comments holds for the $(10,1) \oplus (1,
10)$ chiral multiplet for the same reasons as in the example shown
above.

\subsection{Closure of the Chiral $SU_L(3) \times SU_R(3)$ Algebra
in the $(3,\overline{3}) \oplus (\overline{3}, 3)$ Multiplet}
\label{ssect:closure33a}

The vector charge $Q^a = \int d {\bf x}J_{0}^a (t,{\bf x})$ defined
by the current given in Eq.~(\ref{e:vec1SU(3)33}), together with the
axial charge $Q_{5}^a = \int d {\bf x}J_{05}^a (t,{\bf x})$ defined
by the current given in Eq.~(\ref{e:axi1SU(3)33}) ought to close the
chiral algebra defined by Eqs.~(\ref{e:cc3VVV}), (\ref{e:cc3AVA})
and (\ref{e:cc3AAA}). Eqs.~(\ref{e:cc3VVV}) and (\ref{e:cc3AVA})
hold here, whereas Eq.~(\ref{e:cc3AAA}) is the non-trivial one: the
diagonal $D$ charge of $N$ ($Q_{5D}^a(N)$) axial charge,
\begin{eqnarray}
Q_{5D}^a(N) &=&  ~~~ \int d {\bf x} \, \left(\overline{N} \gamma_{0}
\gamma_{5} \,{\bf D}^{a} \,N \right)
\label{e:offdiag1dSU(3)} \,, \\
Q_{D}^a(N) &=&  ~~~ \int d {\bf x} \, \left(\overline{N} \gamma_{0}
\,{\bf D}^{a} \,N \right) \label{e:offdiag1dNSU(3)} \, ,  \
\end{eqnarray}
lead to
\begin{eqnarray}
\left[Q_{5D}^a(N) , Q_{5D}^b(N) \right] &=& \int d {\bf x}
\left(\overline{N} \gamma_{0} \,\left({\bf D}^{a} {\bf D}^{b} -
{\bf D}^{b} {\bf D}^{a} \right) N \right) \, .
\label{e:ccrAAVS36b} \
\end{eqnarray}
It turns out that the off-diagonal terms in the axial charge
\begin{eqnarray}
Q_{5}^a(N,\Lambda) &=& \int d {\bf x} \, \Bigg( \sqrt{\frac{2}{3}}
\left(\overline{N} \gamma_{0} \gamma_{5} \,{\bf T}^{a \dagger}_{1/8}
\,\Lambda + \overline{\Lambda} \gamma_{0} \gamma_{5} \,{\bf
T}^a_{1/8} \,N \right) \Bigg) ~, \label{e:axioffdiag1Sigma}\
\end{eqnarray}
play a crucial role in the closure of the chiral commutator
Eq.~(\ref{e:cc3AAA}). The additional terms in the commutator add up
to
\begin{eqnarray}
\left[Q_{5}^a(N,\Delta) , Q_{5}^b(N,\Delta) \right] &=&
{\frac{2}{3}} \int d {\bf x}  \overline{N} \gamma_{0} \,\left({\bf
T}^{a\dagger}_{1/8} {\bf T}^{b}_{1/8} - {\bf T}^{b \dagger}_{1/8}
{\bf T}^{a}_{1/8} \right) \,N \, , \label{e:ccrAAVS36a} \
\end{eqnarray}
which provide the ``missing'' factors due to the following
properties of the off-diagonal isospin operators ${\bf T}^{i}_{1/8}$
and ${\bf D}^{i}$ matrices
\begin{eqnarray}
i \,f^{ijk} ({\bf F}_{(8)}^{k}) &=&  ({\bf D}^{i} {\bf D}^{j} - {\bf
D}^{j} {\bf D}^{i}) + {2\over3} ({\bf T}^{i\dagger}_{1/8} {\bf
T}^{j}_{1/8} - {\bf T}^{j\dagger}_{1/8} {\bf T}^{i}_{1/8}) \, .
\label{e:app1bSigma}
\end{eqnarray}
Therefore, the chiral algebra Eqs.~(\ref{e:cc3VVV}),
(\ref{e:cc3AVA}) and (\ref{e:cc3AAA}) close.

\subsection{Closure of the Chiral $SU_L(3) \times SU_R(3)$ Algebra
in the $(3,6) \oplus (6, 3)$ Multiplet} \label{ssect:closure36a}

The vector charge $Q^a = \int d {\bf x}J_{0}^a (t,{\bf x})$ defined
by the current in Eq.~(\ref{e:vec1SU(3)63}), together with the axial
charge $Q_{5}^a = \int d {\bf x}J_{05}^a (t,{\bf x})$ defined by the
current in Eq.~(\ref{e:axi1SU(3)63}) ought to close the chiral
algebra defined by Eqs.~(\ref{e:cc3VVV}), (\ref{e:cc3AVA}) and
(\ref{e:cc3AAA}). Eqs.~(\ref{e:cc3VVV}) and (\ref{e:cc3AVA}) hold
here, whereas Eq.~(\ref{e:cc3AAA}) is once again the non-trivial
one: the fractions $\frac23$ and $\frac13$ in the diagonal $F$
charge of $N$ ($Q_{5}^a(N)$) and $\Delta$ axial charges,
respectively, and the diagonal $D$ charge of $N$ ($Q_{5}^a(N)$):
\begin{eqnarray}
Q_{5F}^a(N) &=& \frac{2}{3} \int d {\bf x} \, \left(\overline{N}
\gamma_{0} \gamma_{5} \,{\bf F}_{(8)}^{a} \,N \right)~
\label{e:offdiag1fSU(3)} \, , \\
Q_{5F}^a(\Delta) &=& \frac{1}{3} \int d {\bf x} \,
\left(\overline{\Delta} \gamma_{0} \gamma_{5} \,{\bf F}_{(10)}^{a}
\,\Delta\right) ~ \label{e:offdiag1DeltaSU(3)} \, , \\
Q_{5D}^a(N) &=&  ~~~ \int d {\bf x} \, \left(\overline{N} \gamma_{0}
\gamma_{5} \,{\bf D}^{a} \,N \right)~ \label{e:offdiag1dSU(3)_2} \, ,
\end{eqnarray}
lead to
\begin{eqnarray}\label{e:ccrAAAdiag1SU(3)}
\left[Q_{5D+F}^a(N) , Q_{5D+F}^b(N) \right] &=& \int d {\bf x}
\Bigg(\overline{N} \gamma_{0} \,\Big(\big({\bf D}^{a} + \frac{2}{3}
{\bf F}_{(8)}^{a}\big) \big({\bf D}^{b} + \frac{2}{3} {\bf
F}_{(8)}^{b}\big) \\ \nonumber && - \big({\bf D}^{b} + \frac{2}{3}
{\bf F}_{(8)}^{b}\big) \big({\bf D}^{a} + \frac{2}{3} {\bf
F}_{(8)}^{a}\big) \Big) N \Bigg) \, , \\ \label{e:ccrAAAdiag2SU(3)A}
\left[Q_{5F}^a(\Delta) , Q_{5F}^b(\Delta) \right] &=& i f^{abc}
\frac19 Q^c(\Delta) \, , \
\end{eqnarray}
lead to "only" one part of the N and Delta vector charges
respectively, on the right-hand side of
Eqs.~(\ref{e:ccrAAAdiag1SU(3)}) and (\ref{e:ccrAAAdiag2SU(3)A}).

Once again, it turns out that the off-diagonal terms in the axial
charge
\begin{eqnarray}
Q_{5}^a(N,\Delta) &=& \int d {\bf x} \, \Bigg( \frac{2}{\sqrt{3}}
\left(\overline{N} \gamma_{0} \gamma_{5} \,{\bf T}^a \,\Delta +
\overline{\Delta} \gamma_{0} \gamma_{5} \,{\bf T}^{a \dagger} \,N
\right) \Bigg)  \label{e:axioffdiag1Sigma2} \, ,
\end{eqnarray}
play a crucial role in the closure of the chiral algebra
Eq.~(\ref{e:cc3AAA}). The additional terms in the commutator add up
to
\begin{eqnarray}
\left[Q_{5}^a(N,\Delta) , Q_{5}^b(N,\Delta) \right] &=&
\frac{4}{3}\int d {\bf x} \left(\overline{N} \gamma_{0}
\,\left({\bf T}^{a} {\bf T}^{b\dagger} - {\bf T}^{b} {\bf T}^{a
\dagger} \right) \,N + \overline{\Delta} \gamma_{0} \,\left({\bf
T}^{a\dagger} {\bf T}^{b} - {\bf T}^{b\dagger} {\bf T}^{a}\right)
\Delta \right) \label{e:ccrAAVS36a2} \, ,
\end{eqnarray}
which provide the ``missing'' factors due to the following
properties of the off-diagonal flavor operators ${\bf T}^{i}$ and
${\bf D}^{i}$ matrices
\begin{eqnarray}
i \,f^{ijk} ({\bf F}_{(8)}^{k})  &=& \Big(\big({\bf D}^{i} +
\frac{2}{3} {\bf F}_{(8)}^{i}\big) \big({\bf D}^{j} + \frac{2}{3}
{\bf F}_{(8)}^{j}\big) - \big({\bf D}^{j} + \frac{2}{3} {\bf
F}_{(8)}^{j}\big) \big({\bf D}^{i} + \frac{2}{3} {\bf
F}_{(8)}^{i}\big)\Big) + {4\over3} ({\bf T}^{i}_{10/8} {\bf
T}^{j\dagger}_{10/8} - {\bf T}^{j}_{10/8} {\bf T}^{i\dagger}_{10/8})
\, , \nonumber \\ i \frac23 \,f^{ijk} {\bf F}_{(10)}^{k} &=& {\bf
T}_{10/8}^{i\dagger} {\bf T}_{10/8}^{j} - {\bf T}_{10/8}^{j\dagger}
{\bf T}_{10/8}^{i} \, . \label{e:app2bSigma}
\end{eqnarray}
Therefore, the chiral algebra Eqs.~(\ref{e:cc3VVV}),
(\ref{e:cc3AVA}) and (\ref{e:cc3AAA}) closes in spite, or perhaps
because of the apparent fractional axial charges ($\frac23$ and
$\frac13$).

\section{Chiral mixing and the axial current}
\label{sect:mix}

A unique feature of the use of the linear chiral representation is
that the axial coupling is determined by the chiral representations,
as given by the coefficients of the axial transformations. For the
nucleon (proton and neutron), chiral representations of $SU_L(2)
\times SU_R(2)$, $(\frac12, 0) (\sim (8,1), (3, \bar 3))$ and $(1,
\frac12) (\sim (6,3))$ provide the nucleon isovector axial coupling
$g_A^{(3)} = 1$ and $5/3$ respectively. Therefore, the mixing of
chiral $(\half,0)$ and $(1,\half)$ nucleons leads to the axial
coupling
\begin{eqnarray}
1.267 &=& g_{A~ (\frac12,0)}^{(1)}~\cos^2\theta + g_{A~
(1,\frac12)}^{(1)}~\sin^2\theta
\nonumber\\
&=& g_{A~ (\frac12,0)}^{(1)}~\cos^2\theta + \frac53~\sin^2\theta \,
, \label{e:axcoupl1}
\end{eqnarray}

\begin{table}
\caption{The Abelian and the non-Abelian axial charges (+ sign
indicates ``naive", - sign ``mirror" transformation properties)
and the non-Abelian chiral multiplets of $J^{P}=\frac12$, Lorentz
representation $(\frac{1}{2},0)$ nucleon and $\Delta$ fields, see
Refs.~\cite{Nagata:2007di,Nagata:2008zzc,Dmitrasinovic:2009vp,Dmitrasinovic:2009vy}.}
\label{tab:spin12b}
\begin{tabular}{llllllll}
\hline \noalign{\smallskip}
case & field & $g_A^{(0)}$ & $g_A^{(1)}$ & $F$ & $D$ & $SU_L(3) \times SU_R(3)$ \\
\noalign{\smallskip}\hline\noalign{\smallskip}
I & $N_1 - N_2$ & $-1$ & $+1$ & $~~0$ & $+1$ & $(3,\overline{3}) \oplus (\overline{3}, 3)$ \\
II & $N_1 + N_2$ & $+3$ & $+1$ & $+1$ & $~~0$ & $(8,1) \oplus (1,8)$ \\
III & $N_1^{'} - N_2^{'}$ & $+1$ & $-1$ & $~~0$ & $-1$ & $(\overline{3},3) \oplus (3,\overline{3})$ \\
IV & $N_1^{'} + N_2^{'}$ & $-3$ & $-1$ & $-1$ & $~~0$ & $(1,8) \oplus (8,1)$ \\
\hline 0 & $\partial_{\mu}(N_3^{\mu} + \frac13 N_4^{\mu})$ & $+1$&
$+\frac53$ & $+\frac23$ & $+1$ & $(6,3) \oplus (3,6)$ \\
\noalign{\smallskip}\hline
\end{tabular}
\end{table}
Three-quark nucleon interpolating fields in QCD have also
well-defined, if perhaps unexpected $U_A(1)$ chiral transformation
properties, see Table \ref{tab:spin12b}, that can be used to predict
the isoscalar axial coupling $g_{A~\rm mix.}^{(0)}$
\begin{eqnarray}
g_{A~\rm mix.}^{(0)} &=& g_{A~ (\frac12,0)}^{(0)}~\cos^2\theta +
g_{A~ (1,\frac12)}^{(0)}~\sin^2\theta
\nonumber \\
&=& g_{A~ (\frac12,0)}^{(0)}~\cos^2\theta + \sin^2\theta ,
\label{e:axcoupl0}
\end{eqnarray}
together with the mixing angle $\theta$ extracted from
Eq.~(\ref{e:axcoupl1}). Note, however, that due to the different
(bare) non-Abelian $g_{A}^{(1)}$ and Abelian $g_{A}^{(0)}$ axial
couplings, see Table \ref{tab:spin12b}, the mixing formulae
Eq.~(\ref{e:axcoupl0}) give substantially different predictions from
one case to another, see Table~\ref{tab:axcoupl1r}.
\begin{table}[tbh]
\begin{center}
\caption{The values of the baryon isoscalar axial coupling
constant predicted from the naive mixing and $g_{A~ \rm
expt.}^{(1)}=1.267$; compare with $g_{A~ \rm expt.}^{(0)}=0.33 \pm
0.03 \pm 0.05$, $F$=$0.459\pm0.008$ and $D$=$0.798\pm0.008$,
leading to $F/D = 0.571 \pm 0.005$,
Ref.~\cite{Yamanishi:2007zza}.}
\begin{tabular}{ccccccccc}
\hline \hline case & ($g_{A}^{(1)}$,$g_{A}^{(0)}$) & $g_{A~\rm
mix.}^{(1)}$ & $\theta_{i}$ & $g_{A~\rm mix.}^{(0)}$ & $g_{A~\rm
mix.}^{(0)}$ & $F$ & $F$/$D$ \\
\hline I & $(+1,-1)$ & $\frac13(4 - \cos 2 \theta)$ &
$39.3^o$ & $- \cos 2 \theta$ & -0.20 & 0.267 & 0.267 \\
II & $(+1,+3)$ & $\frac13(4 - \cos 2 \theta)$ & $39.3^o$
& $(2\cos 2 \theta + 1)$ & 2.20 & 0.866 & 2.16 \\
III & $(-1,+1)$ & $\frac13(1 - 4\cos 2 \theta)$ & $67.2^o$ &
$1$ & 1.00 & 0.567 & 0.81  \\
IV & $(-1,-3)$ & $\frac13(1 - 4\cos 2 \theta)$ & $67.2^o$ &
$-(2\cos 2 \theta + 1)$ & 0.40 & 0.417 & 0.491 \\
 \hline \hline
\end{tabular}
\label{tab:axcoupl1r}
\end{center}
\end{table}
We can see in Table~\ref{tab:axcoupl1r} that the two best candidates
are cases I and IV, with $g_A^{(0)} = - 0.2$ and $g_{A}^{(0)} =
0.4$, respectively, the latter being within the error bars of the
measured value $g_{A~ \rm expt.}^{(0)} = 0.33 \pm 0.08$,
\cite{Bass:2007zzb,Ageev:2007du}. Moreover, this scheme predicts the
$F$ and $D$ values, as well:
\begin{eqnarray}
F &=& F_{(\frac12,0)}~\cos^2\theta +
F_{(1,\frac12)}^{(1)}~\sin^2\theta ,
\nonumber\\
&=& F_{(\frac12,0)}~\cos^2\theta + \frac23~\sin^2\theta \label{e:F} \\
D &=& D_{(\frac12,0)}~\cos^2\theta + D_{(1,\frac12)}~\sin^2\theta
\nonumber \\
&=& D_{(\frac12,0)}~\cos^2\theta + \sin^2\theta , \label{e:D}
\end{eqnarray}
where we have used the $F$ and $D$ values for different chiral
multiplets as listed in Table~\ref{tab:spin12b}.

Cases I and IV, with $F$/$D$ = 0.267 and 0.491, respectively,
ought to be compared with $F$/$D$ = $0.571 \pm 0.005$
\footnote{Note that the Ref.~\cite{Yamanishi:2007zza} values add
up to F+D = $1.312\pm 0.002$, more than 2-$\sigma$ away from the
experimental constraint $\neq 1.269 \pm 0.002$.}. Case I is, of
course, the well-known ``Ioffe current'', which reproduces the
nucleon's properties in QCD lattice and sum rules calculations.
The latter is a ``mirror'' opposite of the orthogonal complement
to the Ioffe current, an interpolating field that, to our
knowledge, has not been used in QCD thus far.

Manifestly, a linear superposition of any three fields (except for
the mixtures of cases II and III, IV above, which yield complex
mixing angles) should give a perfect fit to the central values of
the experimental axial couplings and predict the $F$ and $D$
values. Such a three-field admixture introduces new free
parameters (besides the two already introduced mixing angles, e.g.
$\theta_{1}$ and $\theta_{4}$, we have the relative/mutual mixing
angle $\theta_{14}$, as the two nucleon fields I and IV may also
mix). One may subsume the sum and the difference of the two angles
$\theta_{1}$ and $\theta_{4}$ into the new angle $\theta$, and
define $\varphi \doteq \theta_{14}$ (this relationship depends on
the precise definition of the mixing angles $\theta_{1}$,
$\theta_{4}$ and $\theta_{14}$); thus we find two equations with
two unknowns of the general form:
\begin{align}
\frac{5}{3}\,{\sin}^2 \theta + {\cos}^2\theta\,\left(g_A^{(1)}
{\cos}^2 \varphi + g_A^{(1)\prime}{\sin}^2\varphi \right) &= 1.267 \\
{\sin}^2 \theta + {\cos}^2 \theta\, \left(g_A^{(0)}{\cos}^2
\varphi + g_A^{(0)\prime}\,{\sin}^2 \varphi \right) &= 0.33 \pm
0.08
\end{align}
The solutions to these equations (the values of the mixing angles
$\theta,\varphi$) provide, at the same time, input for the
prediction of $F$ and $D$:
\begin{align}
\cos^2\theta\,\left(F\, {\cos}^2 \varphi +
F^{\prime}\,{\sin}^2\varphi \right) + \frac23~\sin^2\theta
&= F  \label{e:F1} \\
\cos^2\theta\,\left(D\, {\cos}^2 \varphi +
D^{\prime}\,{\sin}^2\varphi \right) + \sin^2\theta &= D.
\label{e:D1}
\end{align}
The values of the mixing angles ($\theta,\varphi$) obtained from
this straightforward fit to the baryon axial coupling constants
are shown in Table \ref{tab:axcoupl2b}.
\begin{table}[tbh]
\begin{center}
\caption{The values of the mixing angles obtained from the simple
fit to the baryon axial coupling constants and the predicted values
of axial $F$ and $D$ couplings. The experimental values are
$F$=$0.459\pm0.008$ and $D$=$0.798\pm0.008$, leading to $F/D = 0.571
\pm 0.005$, Ref. \cite{Yamanishi:2007zza}.}
\begin{tabular}{cccccccc}
\hline \hline case & $g_{A~ \rm expt.}^{(3)}$ & $g_{A~ \rm
expt.}^{(0)}$ & $\theta$ & $\varphi$ & $F$ & $D$ & $F$/$D$ \\
\hline I-II & 1.267 & $0.33$ & $39.3^o$ & $28.0^o \pm 2.3^o$ &
$0.399 \pm 0.02$ & 0.868$\mp 0.02$ &
$0.460\pm 0.04$ \\
I-III & 1.267 & $0.33$ & $50.7^o \pm 1.8^o$ & $23.9^o \pm 2.9^o$ &
$0.399 \pm 0.02$ & 0.868$\mp 0.02$ &
$0.460 \pm 0.04$ \\
I-IV & 1.267 & $0.33$ & $63.2^o \pm 4.0^o$ & $54^o \pm 23^o$ &
$0.399 \pm 0.02$ & 0.868$\mp 0.02$ &
$0.460 \pm 0.04$ \\
\hline \hline
\end{tabular}
\label{tab:axcoupl2b}
\end{center}
\end{table}
Note that all three admissible scenarios (i.e. choices of pairs of
fields admixed to the (6,3) one that lead to real mixing angles)
yield the same values of $F$ and $D$. This is due to the fact that
all three-quark baryon fields satisfy the following relation
$g_{A}^{(0)} = 3F - D = \sqrt{3} g_{A}^{(8)}$~\cite{Jido09}.
The first relation $g_{A}^{(0)} = 3F - D$ was not
expected, as the flavor-singlet properties, such as $g_{A}^{(0)}$
are generally expected to be independent of the flavor-octet ones,
such as $F,D$. Yet, it is not unnatural, either, as it indicates
the absence of polarized $s \bar s$ pairs in these SU(3)
symmetric, three-quark nucleon interpolators.
In order to show that, we define $g_{A}^{(0)} = \Delta u + \Delta
d + \Delta s$ and $g_{A}^{(8)} = {1\over\sqrt3}(\Delta u + \Delta
d - 2 \Delta s )$, where $\Delta q$ are the (corresponding flavor)
quark contributions to the matrix element of the nucleon's axial
vector current $\Delta q = \langle N | \bar q \gamma_\mu \gamma_5
q |N \rangle$. We see that $g_{A}^{(0)} \sim g_{A}^{(8)}$ only if
$\Delta s = 0$.

Thus, the relation $g_{A}^{(0)} = 3F - D $ appears to
depend on the choice of
three-quark interpolating fields as a source of admixed mirror
fields and may well change when one considers other interpolating
fields, such as the five-quark (``pentaquark") ones for
example\footnote{Note, however, that five- and more quark, and
derivative interpolating fields are not the only ones that can
produce mirror fields, however: so can the one-gluon-three-quark
``hybrid baryon" interpolators, which necessarily have the same
chiral properties as the corresponding three-quark fields.}. In
that sense a deviation of the measured values of $g_{A}^{(0)}$ and
$g_{A}^{(8)} = \frac{1}{\sqrt{3}}(3F - D)$ from this relation may
well be seen as a measure of the contribution of higher-order
configurations' to the baryon ground state. It seems very
difficult, however, to evaluate $F$ and $D$ for specific
higher-order configurations without going through the procedure
outlined in Ref. \cite{Chen:2008qv} for the ``pentaquark"
interpolator chiral multiplets \footnote{If one were to assign one
particular source of mirror fields, for example some ``pentaquark"
interpolators, then one could try to determine the contribution of
$s \bar s$ pairs to the flavor singlet axial coupling.}.

Some of the ideas used above have also been used in some of the
following early papers: two-chiral-multiplet mixing was considered
long ago by Harari \cite{Harari:1966yq}, and by Weinberg
\cite{Weinberg:1969hw}, for example. Moreover, special cases of
three-field/configuration chiral mixing have been considered by
Harari \cite{Harari:1966jz}, and by Gerstein and Lee
\cite{Gerstein:1966zz} in the context of the (``collinear") $U(3)
\times U(3)$ current algebra at infinite momentum. One (obvious)
distinction from these early precedents is our use of QCD
interpolating fields, which appeared only in the early 1980's, and
the (perhaps less obvious) issue of baryons' flavor-singlet axial
current (a.k.a. the $U_A(1)$), that was (seriously) raised yet
another decade later. We emphasize here that our results are based
on the $U_L(3) \times U_R(3)$ chiral algebra of space-integrated
time components of currents, without any assumptions about
saturation of this algebra by one-particle states, or its
dependence on any one particular frame of reference. Indeed, our
nucleon interpolating fields transform as the
$(\frac12,0)+(0,\frac12)$ representation of the Lorentz group,
thus making the Noether currents (fully) Lorentz covariant, so
that our results hold in any frame.

\section{Summary and Outlook}
\label{sect:summary}

We have re-organized the results of our previous paper
\cite{Chen:2008qv} into the (perhaps more) conventional form for the
baryon octet using $F$ and $D$ coupling ($SU(3)$ structure
constants). This means that, {\it inter alia}, we have explicitly
written down (perhaps for the first time) the chiral transformations
of the $(6,3)\oplus(3,6)$ octet and decimet fields in the $SU(3)$
particle (octet and decimet) basis.

In the process we have independently constructed $SU(3)$ generators
of the decimet and derived a set of $SU(3)$ Clebsch-Gordan
coefficients in the ``natural'' convention, which means that all
isospin $SU(2)$ sub-multiplets of the octet and the decimet have
standard isospin $SU(2)$ generators.

Then we used the above mentioned $SU(3)$ Clebsch-Gordan coefficients
to explicitly check the closure of the $SU_L(3) \times SU_R(3)$
chiral algebra in the $SU(3)$ particle basis, which forms an
independent check/confirmation of the calculation.

Next, we investigated the phenomenological consequence for the
baryon axial currents, of the chiral $[(6,3)\oplus(3,6)]$ multiplet
mixing with other three-quark baryon field multiplets, such as the
$[(3,\overline{3}) \oplus (\overline{3}, 3)]$ and $[(8,1)\oplus
(1,8)]$. The results of the three-field (``two-angle'') mixing are
interesting: all permissible combinations fields lead to the same
$F$/$D$ prediction, that is in reasonable agreement with experiment.
This identity of results is a consequence of the relation
$g_{A}^{(0)}= 3F-D$ between the flavor singlet axial coupling
$g_{A}^{(0)}$ and the (previously unrelated) flavor octet $F$ and
$D$ values. That relation is a unique feature of the three-quark
interpolating fields and any potential departures from it may be
attributed to fields with a number of quarks higher than three.

The next step, left for the future, is to investigate $SU_L(3)
\times SU_R(3)$ chiral invariant interactions and the $SU(3)\times
SU(3) \to SU(2)\times SU(2)$ symmetry breaking/reduction and to the
study of the chiral $SU(2)\times SU(2)$ properties of hyperons. Then
one may consider explicit chiral symmetry breaking corrections to
the axial and the vector currents, which are related to the
$SU(3)\times SU(3)$ symmetry breaking meson-nucleon derivative
interactions, not just the explicit $SU(3)$ symmetry breaking ones
that have been considered thus far (see
Ref.~\cite{Yamanishi:2007zza} and the previous subsection, above).

\section*{Acknowledgments}
\label{ack}

We wish to thank Profs. Daisuke Jido, Akira Ohnishi and Makoto Oka
for valuable conversations regarding the present work. One of us
(V.D.) wishes to thank the RCNP, Osaka University, under whose
auspices this work was begun, and the Yukawa Institute for
Theoretical Physics, Kyoto, (molecular workshop ``Algebraic aspects
of chiral symmetry for the study of excited baryons") where it was
finished, for kind hospitality and financial support.

\appendix
\section{SU(3) Octet, Decimet Generators and 8x10 Transition Matrices}
\label{sect:Octet_Decimet_Generators Veljko}

\subsection{Octet ``Generator" 8x8 Matrices ${\rm {\bf D}^{a}},~ {\rm {\bf
F}_{(8)}^{a}}$ in the Particle Basis} \label{sect:Octet Generators
Veljko}

\begin{eqnarray}
{\rm ({\bf D}^{1} + {2\over3} {\bf F}_{(8)}^{1})} &=& \left(
\begin{array}{llllllll|l}
 0 & \frac{5}{6} & 0 & 0 & 0 & 0 & 0 & 0 & p \\
 \frac{5}{6} & 0 & 0 & 0 & 0 & 0 & 0 & 0 & n \\
 0 & 0 & 0 & \frac{\sqrt{2}}{3} & 0 & 0 & 0 & -\frac{1}{\sqrt{6}}
 & \Sigma^{+}  \\
 0 & 0 & \frac{\sqrt{2}}{3} & 0 & \frac{\sqrt{2}}{3} & 0 & 0 & 0
 & \Sigma^{0}\\
 0 & 0 & 0 & \frac{\sqrt{2}}{3} & 0 & 0 & 0 & \frac{1}{\sqrt{6}}
 & \Sigma^{-} \\
 0 & 0 & 0 & 0 & 0 & 0 & -\frac{1}{6} & 0 & \Xi^{0} \\
 0 & 0 & 0 & 0 & 0 & -\frac{1}{6} & 0 & 0 & \Xi^{-} \\
 0 & 0 & -\frac{1}{\sqrt{6}} & 0 & \frac{1}{\sqrt{6}} & 0 & 0 & 0 & \Lambda_8 \\
 \hline \\
 p & n & \Sigma^{+} & \Sigma^{0} & \Sigma^{-} & \Xi^{0} & \Xi^{-} &
 \Lambda_8
\end{array}
\right) \\
{\rm {\bf D}^{1}} &=& \left(
\begin{array}{llllllll}
 0 & \frac{1}{2} & 0 & 0 & 0 & 0 & 0 & 0 \\
 \frac{1}{2} & 0 & 0 & 0 & 0 & 0 & 0 & 0 \\
 0 & 0 & 0 & 0 & 0 & 0 & 0 & -\frac{1}{\sqrt{6}} \\
 0 & 0 & 0 & 0 & 0 & 0 & 0 & 0 \\
 0 & 0 & 0 & 0 & 0 & 0 & 0 & \frac{1}{\sqrt{6}} \\
 0 & 0 & 0 & 0 & 0 & 0 & -\frac{1}{2} & 0 \\
 0 & 0 & 0 & 0 & 0 & -\frac{1}{2} & 0 & 0 \\
 0 & 0 & -\frac{1}{\sqrt{6}} & 0 & \frac{1}{\sqrt{6}} & 0 & 0 & 0
\end{array}
\right)  \\
{\bf F}_{(8)}^{1} &=&  \left(
\begin{array}{llllllll}
 0 & \frac{1}{2} & 0 & 0 & 0 & 0 & 0 & 0 \\
 \frac{1}{2} & 0 & 0 & 0 & 0 & 0 & 0 & 0 \\
 0 & 0 & 0 & \frac{1}{\sqrt{2}} & 0 & 0 & 0 & 0 \\
 0 & 0 & \frac{1}{\sqrt{2}} & 0 & \frac{1}{\sqrt{2}} & 0 & 0 & 0 \\
 0 & 0 & 0 & \frac{1}{\sqrt{2}} & 0 & 0 & 0 & 0 \\
 0 & 0 & 0 & 0 & 0 & 0 & \frac{1}{2} & 0 \\
 0 & 0 & 0 & 0 & 0 & \frac{1}{2} & 0 & 0 \\
 0 & 0 & 0 & 0 & 0 & 0 & 0 & 0
\end{array}
\right)
\end{eqnarray}

\begin{eqnarray}
{\rm ({\bf D}^{2} + {2\over3} {\bf F}_{(8)}^{2})} &=& \left(
\begin{array}{llllllll}
 0 & -\frac{5 i}{6} & 0 & 0 & 0 & 0 & 0 & 0 \\
 \frac{5 i}{6} & 0 & 0 & 0 & 0 & 0 & 0 & 0 \\
 0 & 0 & 0 & -\frac{i \sqrt{2}}{3} & 0 & 0 & 0 &
   \frac{i}{\sqrt{6}} \\
 0 & 0 & \frac{i \sqrt{2}}{3} & 0 & -\frac{i \sqrt{2}}{3} & 0 & 0
   & 0 \\
 0 & 0 & 0 & \frac{i \sqrt{2}}{3} & 0 & 0 & 0 & \frac{i}{\sqrt{6}}
   \\
 0 & 0 & 0 & 0 & 0 & 0 & \frac{i}{6} & 0 \\
 0 & 0 & 0 & 0 & 0 & -\frac{i}{6} & 0 & 0 \\
 0 & 0 & -\frac{i}{\sqrt{6}} & 0 & -\frac{i}{\sqrt{6}} & 0 & 0 & 0
\end{array}
\right) \\
{\rm {\bf D}^{2}} &=& \left(
\begin{array}{llllllll}
 0 & -\frac{i}{2} & 0 & 0 & 0 & 0 & 0 & 0 \\
 \frac{i}{2} & 0 & 0 & 0 & 0 & 0 & 0 & 0 \\
 0 & 0 & 0 & 0 & 0 & 0 & 0 & \frac{i}{\sqrt{6}} \\
 0 & 0 & 0 & 0 & 0 & 0 & 0 & 0 \\
 0 & 0 & 0 & 0 & 0 & 0 & 0 & \frac{i}{\sqrt{6}} \\
 0 & 0 & 0 & 0 & 0 & 0 & \frac{i}{2} & 0 \\
 0 & 0 & 0 & 0 & 0 & -\frac{i}{2} & 0 & 0 \\
 0 & 0 & -\frac{i}{\sqrt{6}} & 0 & -\frac{i}{\sqrt{6}} & 0 & 0 & 0
\end{array}
\right) \\
{\bf F}_{(8)}^{2} &=& \left(
\begin{array}{llllllll}
 0 & -\frac{i}{2} & 0 & 0 & 0 & 0 & 0 & 0 \\
 \frac{i}{2} & 0 & 0 & 0 & 0 & 0 & 0 & 0 \\
 0 & 0 & 0 & -\frac{i}{\sqrt{2}} & 0 & 0 & 0 & 0 \\
 0 & 0 & \frac{i}{\sqrt{2}} & 0 & -\frac{i}{\sqrt{2}} & 0 & 0 & 0 \\
 0 & 0 & 0 & \frac{i}{\sqrt{2}} & 0 & 0 & 0 & 0 \\
 0 & 0 & 0 & 0 & 0 & 0 & -\frac{i}{2} & 0 \\
 0 & 0 & 0 & 0 & 0 & \frac{i}{2} & 0 & 0 \\
 0 & 0 & 0 & 0 & 0 & 0 & 0 & 0
\end{array}
\right)
\end{eqnarray}

\begin{eqnarray}
{\rm ({\bf D}^{3} + {2\over3} {\bf F}_{(8)}^{3})} &=& \left(
\begin{array}{llllllll}
 \frac{5}{6} & 0 & 0 & 0 & 0 & 0 & 0 & 0 \\
 0 & -\frac{5}{6} & 0 & 0 & 0 & 0 & 0 & 0 \\
 0 & 0 & \frac{2}{3} & 0 & 0 & 0 & 0 & 0 \\
 0 & 0 & 0 & 0 & 0 & 0 & 0 & \frac{1}{\sqrt{3}} \\
 0 & 0 & 0 & 0 & -\frac{2}{3} & 0 & 0 & 0 \\
 0 & 0 & 0 & 0 & 0 & -\frac{1}{6} & 0 & 0 \\
 0 & 0 & 0 & 0 & 0 & 0 & \frac{1}{6} & 0 \\
 0 & 0 & 0 & \frac{1}{\sqrt{3}} & 0 & 0 & 0 & 0
\end{array}
\right) \\
{\rm {\bf D}^{3}} &=& \left(
\begin{array}{llllllll}
 \frac{1}{2} & 0 & 0 & 0 & 0 & 0 & 0 & 0 \\
 0 & -\frac{1}{2} & 0 & 0 & 0 & 0 & 0 & 0 \\
 0 & 0 & 0 & 0 & 0 & 0 & 0 & 0 \\
 0 & 0 & 0 & 0 & 0 & 0 & 0 & \frac{1}{\sqrt{3}} \\
 0 & 0 & 0 & 0 & 0 & 0 & 0 & 0 \\
 0 & 0 & 0 & 0 & 0 & -\frac{1}{2} & 0 & 0 \\
 0 & 0 & 0 & 0 & 0 & 0 & \frac{1}{2} & 0 \\
 0 & 0 & 0 & \frac{1}{\sqrt{3}} & 0 & 0 & 0 & 0
\end{array}
\right) \\
{\bf F}_{(8)}^{3} &=& \left(
\begin{array}{llllllll}
 \frac{1}{2} & 0 & 0 & 0 & 0 & 0 & 0 & 0 \\
 0 & -\frac{1}{2} & 0 & 0 & 0 & 0 & 0 & 0 \\
 0 & 0 & 1 & 0 & 0 & 0 & 0 & 0 \\
 0 & 0 & 0 & 0 & 0 & 0 & 0 & 0 \\
 0 & 0 & 0 & 0 & -1 & 0 & 0 & 0 \\
 0 & 0 & 0 & 0 & 0 & \frac{1}{2} & 0 & 0 \\
 0 & 0 & 0 & 0 & 0 & 0 & -\frac{1}{2} & 0 \\
 0 & 0 & 0 & 0 & 0 & 0 & 0 & 0
\end{array}
\right)
\end{eqnarray}

\begin{eqnarray}
{\rm ({\bf D}^{4} + {2\over3} {\bf F}_{(8)}^{4})} &=& \left(
\begin{array}{llllllll}
 0 & 0 & 0 & \frac{1}{6 \sqrt{2}} & 0 & 0 & 0 &
   -\frac{\sqrt{\frac{3}{2}}}{2} \\
 0 & 0 & 0 & 0 & \frac{1}{6} & 0 & 0 & 0 \\
 0 & 0 & 0 & 0 & 0 & -\frac{5}{6} & 0 & 0 \\
 \frac{1}{6 \sqrt{2}} & 0 & 0 & 0 & 0 & 0 & -\frac{5}{6 \sqrt{2}}
   & 0 \\
 0 & \frac{1}{6} & 0 & 0 & 0 & 0 & 0 & 0 \\
 0 & 0 & -\frac{5}{6} & 0 & 0 & 0 & 0 & 0 \\
 0 & 0 & 0 & -\frac{5}{6 \sqrt{2}} & 0 & 0 & 0 & -\frac{1}{2
   \sqrt{6}} \\
 -\frac{\sqrt{\frac{3}{2}}}{2} & 0 & 0 & 0 & 0 & 0 & -\frac{1}{2
   \sqrt{6}} & 0
\end{array}
\right) \\
{\rm {\bf D}^{4}} &=& \left(
\begin{array}{llllllll}
 0 & 0 & 0 & \frac{1}{2 \sqrt{2}} & 0 & 0 & 0 & -\frac{1}{2
   \sqrt{6}} \\
 0 & 0 & 0 & 0 & \frac{1}{2} & 0 & 0 & 0 \\
 0 & 0 & 0 & 0 & 0 & -\frac{1}{2} & 0 & 0 \\
 \frac{1}{2 \sqrt{2}} & 0 & 0 & 0 & 0 & 0 & -\frac{1}{2 \sqrt{2}}
   & 0 \\
 0 & \frac{1}{2} & 0 & 0 & 0 & 0 & 0 & 0 \\
 0 & 0 & -\frac{1}{2} & 0 & 0 & 0 & 0 & 0 \\
 0 & 0 & 0 & -\frac{1}{2 \sqrt{2}} & 0 & 0 & 0 & \frac{1}{2
   \sqrt{6}} \\
 -\frac{1}{2 \sqrt{6}} & 0 & 0 & 0 & 0 & 0 & \frac{1}{2 \sqrt{6}}
   & 0
\end{array}
\right) \\
{\bf F}_{(8)}^{4} &=& \left(
\begin{array}{llllllll}
 0 & 0 & 0 & -\frac{1}{2 \sqrt{2}} & 0 & 0 & 0 &
   -\frac{\sqrt{\frac{3}{2}}}{2} \\
 0 & 0 & 0 & 0 & -\frac{1}{2} & 0 & 0 & 0 \\
 0 & 0 & 0 & 0 & 0 & -\frac{1}{2} & 0 & 0 \\
 -\frac{1}{2 \sqrt{2}} & 0 & 0 & 0 & 0 & 0 & -\frac{1}{2 \sqrt{2}} & 0
   \\
 0 & -\frac{1}{2} & 0 & 0 & 0 & 0 & 0 & 0 \\
 0 & 0 & -\frac{1}{2} & 0 & 0 & 0 & 0 & 0 \\
 0 & 0 & 0 & -\frac{1}{2 \sqrt{2}} & 0 & 0 & 0 &
   -\frac{\sqrt{\frac{3}{2}}}{2} \\
 -\frac{\sqrt{\frac{3}{2}}}{2} & 0 & 0 & 0 & 0 & 0 &
   -\frac{\sqrt{\frac{3}{2}}}{2} & 0
\end{array}\right)
\end{eqnarray}

\begin{eqnarray}
{\rm ({\bf D}^{5} + {2\over3} {\bf F}_{(8)}^{5})} &=& \left(
\begin{array}{llllllll}
 0 & 0 & 0 & -\frac{i}{6 \sqrt{2}} & 0 & 0 & 0 & \frac{1}{2} i
   \sqrt{\frac{3}{2}} \\
 0 & 0 & 0 & 0 & -\frac{i}{6} & 0 & 0 & 0 \\
 0 & 0 & 0 & 0 & 0 & \frac{5 i}{6} & 0 & 0 \\
 \frac{i}{6 \sqrt{2}} & 0 & 0 & 0 & 0 & 0 & \frac{5 i}{6 \sqrt{2}}
   & 0 \\
 0 & \frac{i}{6} & 0 & 0 & 0 & 0 & 0 & 0 \\
 0 & 0 & -\frac{5 i}{6} & 0 & 0 & 0 & 0 & 0 \\
 0 & 0 & 0 & -\frac{5 i}{6 \sqrt{2}} & 0 & 0 & 0 & -\frac{i}{2
   \sqrt{6}} \\
 -\frac{1}{2} i \sqrt{\frac{3}{2}} & 0 & 0 & 0 & 0 & 0 &
   \frac{i}{2 \sqrt{6}} & 0
\end{array}
\right) \\
{\rm {\bf D}^{5}} &=& \left(
\begin{array}{llllllll}
 0 & 0 & 0 & -\frac{i}{2 \sqrt{2}} & 0 & 0 & 0 & \frac{i}{2
   \sqrt{6}} \\
 0 & 0 & 0 & 0 & -\frac{i}{2} & 0 & 0 & 0 \\
 0 & 0 & 0 & 0 & 0 & \frac{i}{2} & 0 & 0 \\
 \frac{i}{2 \sqrt{2}} & 0 & 0 & 0 & 0 & 0 & \frac{i}{2 \sqrt{2}} &
   0 \\
 0 & \frac{i}{2} & 0 & 0 & 0 & 0 & 0 & 0 \\
 0 & 0 & -\frac{i}{2} & 0 & 0 & 0 & 0 & 0 \\
 0 & 0 & 0 & -\frac{i}{2 \sqrt{2}} & 0 & 0 & 0 & \frac{i}{2
   \sqrt{6}} \\
 -\frac{i}{2 \sqrt{6}} & 0 & 0 & 0 & 0 & 0 & -\frac{i}{2 \sqrt{6}}
   & 0
\end{array}
\right) \\
{\bf F}_{(8)}^{5} &=& \left(
\begin{array}{llllllll}
 0 & 0 & 0 & \frac{i}{2 \sqrt{2}} & 0 & 0 & 0 & \frac{1}{2} i
   \sqrt{\frac{3}{2}} \\
 0 & 0 & 0 & 0 & \frac{i}{2} & 0 & 0 & 0 \\
 0 & 0 & 0 & 0 & 0 & \frac{i}{2} & 0 & 0 \\
 -\frac{i}{2 \sqrt{2}} & 0 & 0 & 0 & 0 & 0 & \frac{i}{2 \sqrt{2}} & 0
   \\
 0 & -\frac{i}{2} & 0 & 0 & 0 & 0 & 0 & 0 \\
 0 & 0 & -\frac{i}{2} & 0 & 0 & 0 & 0 & 0 \\
 0 & 0 & 0 & -\frac{i}{2 \sqrt{2}} & 0 & 0 & 0 & -\frac{1}{2} i
   \sqrt{\frac{3}{2}} \\
 -\frac{1}{2} i \sqrt{\frac{3}{2}} & 0 & 0 & 0 & 0 & 0 & \frac{1}{2} i
   \sqrt{\frac{3}{2}} & 0
\end{array}
\right)
\end{eqnarray}

\begin{eqnarray}
{\rm ({\bf D}^{6} + {2\over3} {\bf F}_{(8)}^{6})} &=& \left(
\begin{array}{llllllll}
 0 & 0 & -\frac{1}{6} & 0 & 0 & 0 & 0 & 0 \\
 0 & 0 & 0 & -\frac{1}{6 \sqrt{2}} & 0 & 0 & 0 &
   -\frac{\sqrt{\frac{3}{2}}}{2} \\
 -\frac{1}{6} & 0 & 0 & 0 & 0 & 0 & 0 & 0 \\
 0 & -\frac{1}{6 \sqrt{2}} & 0 & 0 & 0 & -\frac{5}{6 \sqrt{2}} & 0
   & 0 \\
 0 & 0 & 0 & 0 & 0 & 0 & -\frac{5}{6} & 0 \\
 0 & 0 & 0 & -\frac{5}{6 \sqrt{2}} & 0 & 0 & 0 & \frac{1}{2
   \sqrt{6}} \\
 0 & 0 & 0 & 0 & -\frac{5}{6} & 0 & 0 & 0 \\
 0 & -\frac{\sqrt{\frac{3}{2}}}{2} & 0 & 0 & 0 & \frac{1}{2
   \sqrt{6}} & 0 & 0
\end{array}
\right) \\
{\rm {\bf D}^{6}} &=& \left(
\begin{array}{llllllll}
 0 & 0 & -\frac{1}{2} & 0 & 0 & 0 & 0 & 0 \\
 0 & 0 & 0 & -\frac{1}{2 \sqrt{2}} & 0 & 0 & 0 & -\frac{1}{2
   \sqrt{6}} \\
 -\frac{1}{2} & 0 & 0 & 0 & 0 & 0 & 0 & 0 \\
 0 & -\frac{1}{2 \sqrt{2}} & 0 & 0 & 0 & -\frac{1}{2 \sqrt{2}} & 0
   & 0 \\
 0 & 0 & 0 & 0 & 0 & 0 & -\frac{1}{2} & 0 \\
 0 & 0 & 0 & -\frac{1}{2 \sqrt{2}} & 0 & 0 & 0 & -\frac{1}{2
   \sqrt{6}} \\
 0 & 0 & 0 & 0 & -\frac{1}{2} & 0 & 0 & 0 \\
 0 & -\frac{1}{2 \sqrt{6}} & 0 & 0 & 0 & -\frac{1}{2 \sqrt{6}} & 0
   & 0
\end{array}
\right) \\
{\bf F}_{(8)}^{6} &=& \left(
\begin{array}{llllllll}
 0 & 0 & \frac{1}{2} & 0 & 0 & 0 & 0 & 0 \\
 0 & 0 & 0 & \frac{1}{2 \sqrt{2}} & 0 & 0 & 0 &
   -\frac{\sqrt{\frac{3}{2}}}{2} \\
 \frac{1}{2} & 0 & 0 & 0 & 0 & 0 & 0 & 0 \\
 0 & \frac{1}{2 \sqrt{2}} & 0 & 0 & 0 & -\frac{1}{2 \sqrt{2}} & 0 & 0
   \\
 0 & 0 & 0 & 0 & 0 & 0 & -\frac{1}{2} & 0 \\
 0 & 0 & 0 & -\frac{1}{2 \sqrt{2}} & 0 & 0 & 0 &
   \frac{\sqrt{\frac{3}{2}}}{2} \\
 0 & 0 & 0 & 0 & -\frac{1}{2} & 0 & 0 & 0 \\
 0 & -\frac{\sqrt{\frac{3}{2}}}{2} & 0 & 0 & 0 &
   \frac{\sqrt{\frac{3}{2}}}{2} & 0 & 0
\end{array}
\right)
\end{eqnarray}

\begin{eqnarray}
{\rm ({\bf D}^{7} + {2\over3} {\bf F}_{(8)}^{7})} &=& \left(
\begin{array}{llllllll}
 0 & 0 & \frac{i}{6} & 0 & 0 & 0 & 0 & 0 \\
 0 & 0 & 0 & \frac{i}{6 \sqrt{2}} & 0 & 0 & 0 & \frac{1}{2} i
   \sqrt{\frac{3}{2}} \\
 -\frac{i}{6} & 0 & 0 & 0 & 0 & 0 & 0 & 0 \\
 0 & -\frac{i}{6 \sqrt{2}} & 0 & 0 & 0 & \frac{5 i}{6 \sqrt{2}} &
   0 & 0 \\
 0 & 0 & 0 & 0 & 0 & 0 & \frac{5 i}{6} & 0 \\
 0 & 0 & 0 & -\frac{5 i}{6 \sqrt{2}} & 0 & 0 & 0 & \frac{i}{2
   \sqrt{6}} \\
 0 & 0 & 0 & 0 & -\frac{5 i}{6} & 0 & 0 & 0 \\
 0 & -\frac{1}{2} i \sqrt{\frac{3}{2}} & 0 & 0 & 0 & -\frac{i}{2
   \sqrt{6}} & 0 & 0
\end{array}
\right) \\
{\rm {\bf D}^{7}} &=& \left(
\begin{array}{llllllll}
 0 & 0 & \frac{i}{2} & 0 & 0 & 0 & 0 & 0 \\
 0 & 0 & 0 & \frac{i}{2 \sqrt{2}} & 0 & 0 & 0 & \frac{i}{2
   \sqrt{6}} \\
 -\frac{i}{2} & 0 & 0 & 0 & 0 & 0 & 0 & 0 \\
 0 & -\frac{i}{2 \sqrt{2}} & 0 & 0 & 0 & \frac{i}{2 \sqrt{2}} & 0
   & 0 \\
 0 & 0 & 0 & 0 & 0 & 0 & \frac{i}{2} & 0 \\
 0 & 0 & 0 & -\frac{i}{2 \sqrt{2}} & 0 & 0 & 0 & -\frac{i}{2
   \sqrt{6}} \\
 0 & 0 & 0 & 0 & -\frac{i}{2} & 0 & 0 & 0 \\
 0 & -\frac{i}{2 \sqrt{6}} & 0 & 0 & 0 & \frac{i}{2 \sqrt{6}} & 0
   & 0
\end{array}
\right) \\
{\bf F}_{(8)}^{7} &=& \left(
\begin{array}{llllllll}
 0 & 0 & -\frac{i}{2} & 0 & 0 & 0 & 0 & 0 \\
 0 & 0 & 0 & -\frac{i}{2 \sqrt{2}} & 0 & 0 & 0 & \frac{1}{2} i
   \sqrt{\frac{3}{2}} \\
 \frac{i}{2} & 0 & 0 & 0 & 0 & 0 & 0 & 0 \\
 0 & \frac{i}{2 \sqrt{2}} & 0 & 0 & 0 & \frac{i}{2 \sqrt{2}} & 0 & 0 \\
 0 & 0 & 0 & 0 & 0 & 0 & \frac{i}{2} & 0 \\
 0 & 0 & 0 & -\frac{i}{2 \sqrt{2}} & 0 & 0 & 0 & \frac{1}{2} i
   \sqrt{\frac{3}{2}} \\
 0 & 0 & 0 & 0 & -\frac{i}{2} & 0 & 0 & 0 \\
 0 & -\frac{1}{2} i \sqrt{\frac{3}{2}} & 0 & 0 & 0 & -\frac{1}{2} i
   \sqrt{\frac{3}{2}} & 0 & 0
\end{array}
\right)
\end{eqnarray}

\begin{eqnarray}
{\rm ({\bf D}^{8} + {2\over3} {\bf F}_{(8)}^{2})} &=& \left(
\begin{array}{llllllll}
 \frac{1}{2 \sqrt{3}} & 0 & 0 & 0 & 0 & 0 & 0 & 0 \\
 0 & \frac{1}{2 \sqrt{3}} & 0 & 0 & 0 & 0 & 0 & 0 \\
 0 & 0 & \frac{1}{\sqrt{3}} & 0 & 0 & 0 & 0 & 0 \\
 0 & 0 & 0 & \frac{1}{\sqrt{3}} & 0 & 0 & 0 & 0 \\
 0 & 0 & 0 & 0 & \frac{1}{\sqrt{3}} & 0 & 0 & 0 \\
 0 & 0 & 0 & 0 & 0 & -\frac{\sqrt{3}}{2} & 0 & 0 \\
 0 & 0 & 0 & 0 & 0 & 0 & -\frac{\sqrt{3}}{2} & 0 \\
 0 & 0 & 0 & 0 & 0 & 0 & 0 & -\frac{1}{\sqrt{3}}
\end{array}
\right) \\
{\rm {\bf D}^{8}} &=& \left(
\begin{array}{llllllll}
 -\frac{1}{2 \sqrt{3}} & 0 & 0 & 0 & 0 & 0 & 0 & 0 \\
 0 & -\frac{1}{2 \sqrt{3}} & 0 & 0 & 0 & 0 & 0 & 0 \\
 0 & 0 & \frac{1}{\sqrt{3}} & 0 & 0 & 0 & 0 & 0 \\
 0 & 0 & 0 & \frac{1}{\sqrt{3}} & 0 & 0 & 0 & 0 \\
 0 & 0 & 0 & 0 & \frac{1}{\sqrt{3}} & 0 & 0 & 0 \\
 0 & 0 & 0 & 0 & 0 & -\frac{1}{2 \sqrt{3}} & 0 & 0 \\
 0 & 0 & 0 & 0 & 0 & 0 & -\frac{1}{2 \sqrt{3}} & 0 \\
 0 & 0 & 0 & 0 & 0 & 0 & 0 & -\frac{1}{\sqrt{3}}
\end{array}
\right) \\
{\bf F}_{(8)}^{8} &=& \left(
\begin{array}{llllllll}
 \frac{\sqrt{3}}{2} & 0 & 0 & 0 & 0 & 0 & 0 & 0 \\
 0 & \frac{\sqrt{3}}{2} & 0 & 0 & 0 & 0 & 0 & 0 \\
 0 & 0 & 0 & 0 & 0 & 0 & 0 & 0 \\
 0 & 0 & 0 & 0 & 0 & 0 & 0 & 0 \\
 0 & 0 & 0 & 0 & 0 & 0 & 0 & 0 \\
 0 & 0 & 0 & 0 & 0 & -\frac{\sqrt{3}}{2} & 0 & 0 \\
 0 & 0 & 0 & 0 & 0 & 0 & -\frac{\sqrt{3}}{2} & 0 \\
 0 & 0 & 0 & 0 & 0 & 0 & 0 & 0
\end{array}
\right)
\end{eqnarray}

\subsection{Octet-Decimet 8x10 Transition Matrices $T^a$}
\label{s:Veljko8x10}

\begin{eqnarray}
T_1 = \left(
\begin{array}{llllllllll|l}
-\frac{1}{\sqrt{2}} & 0 & \frac{1}{\sqrt{6}} & 0 & 0 & 0 & 0 & 0 & 0 & 0 & p \\
0 & -\frac{1}{\sqrt{6}} & 0 & \frac{1}{\sqrt{2}} & 0 & 0 & 0 & 0 & 0 & 0 & n \\
0 & 0 & 0 & 0 & 0 & \frac{1}{2 \sqrt{3}} & 0 & 0 & 0 & 0 & \Sigma^{+} \\
0 & 0 & 0 & 0 & \frac{1}{2 \sqrt{3}} & 0 & \frac{1}{2 \sqrt{3}} & 0 & 0 & 0 & \Sigma^{0}\\
0 & 0 & 0 & 0 & 0 & \frac{1}{2 \sqrt{3}} & 0 & 0 & 0 & 0 & \Sigma^{-} \\
0 & 0 & 0 & 0 & 0 & 0 & 0 & 0 & -\frac{1}{\sqrt{6}} & 0 & \Xi^{0} \\
0 & 0 & 0 & 0 & 0 & 0 & 0 & -\frac{1}{\sqrt{6}} & 0 & 0 & \Xi^{-} \\
0 & 0 & 0 & 0 & \frac{1}{2} & 0 & -\frac{1}{2} & 0 & 0 & 0 & \Lambda_8 \\
\hline \\
\Delta^{++} & \Delta^+ & \Delta^0 & \Delta^- &
\Sigma^{*+} & \Sigma^{*0} & \Sigma^{*-} & \Xi^{*0} & \Xi^{*-} &
\Omega
\end{array}
\right)
\end{eqnarray}

\begin{eqnarray}
T_2 = \left(
\begin{array}{llllllllll}
-\frac{i}{\sqrt{2}} & 0 & -\frac{i}{\sqrt{6}} & 0 & 0 & 0 & 0 & 0 &
0
& 0 \\
0 & -\frac{i}{\sqrt{6}} & 0 & -\frac{i}{\sqrt{2}} & 0 & 0 & 0 & 0 &
0
& 0 \\
0 & 0 & 0 & 0 & 0 & -\frac{i}{2 \sqrt{3}} & 0 & 0 & 0 & 0 \\
0 & 0 & 0 & 0 & \frac{i}{2 \sqrt{3}} & 0 & -\frac{i}{2 \sqrt{3}} & 0
& 0 & 0 \\
0 & 0 & 0 & 0 & 0 & \frac{i}{2 \sqrt{3}} & 0 & 0 & 0 & 0 \\
0 & 0 & 0 & 0 & 0 & 0 & 0 & 0 & \frac{i}{\sqrt{6}} & 0 \\
0 & 0 & 0 & 0 & 0 & 0 & 0 & -\frac{i}{\sqrt{6}} & 0 & 0 \\
0 & 0 & 0 & 0 & \frac{i}{2} & 0 & \frac{i}{2} & 0 & 0 & 0
\end{array}
\right)
\end{eqnarray}

\begin{eqnarray}
T_3 = \left(
\begin{array}{llllllllll}
0 & \sqrt{\frac{2}{3}} & 0 & 0 & 0 & 0 & 0 & 0 & 0 & 0 \\
0 & 0 & \sqrt{\frac{2}{3}} & 0 & 0 & 0 & 0 & 0 & 0 & 0 \\
0 & 0 & 0 & 0 & \frac{1}{\sqrt{6}} & 0 & 0 & 0 & 0 & 0 \\
0 & 0 & 0 & 0 & 0 & 0 & 0 & 0 & 0 & 0 \\
0 & 0 & 0 & 0 & 0 & 0 & -\frac{1}{\sqrt{6}} & 0 & 0 & 0 \\
0 & 0 & 0 & 0 & 0 & 0 & 0 & -\frac{1}{\sqrt{6}} & 0 & 0 \\
0 & 0 & 0 & 0 & 0 & 0 & 0 & 0 & \frac{1}{\sqrt{6}} & 0 \\
0 & 0 & 0 & 0 & 0 & -\frac{1}{\sqrt{2}} & 0 & 0 & 0 & 0
\end{array}
\right)
\end{eqnarray}

\begin{eqnarray}
T_4 = \left(
\begin{array}{llllllllll}
0 & 0 & 0 & 0 & 0 & \frac{1}{2 \sqrt{3}} & 0 & 0 & 0 & 0 \\
0 & 0 & 0 & 0 & 0 & 0 & \frac{1}{\sqrt{6}} & 0 & 0 & 0 \\
-\frac{1}{\sqrt{2}} & 0 & 0 & 0 & 0 & 0 & 0 & \frac{1}{\sqrt{6}} & 0
& 0 \\
0 & -\frac{1}{\sqrt{3}} & 0 & 0 & 0 & 0 & 0 & 0 & \frac{1}{2
\sqrt{3}} & 0 \\
0 & 0 & -\frac{1}{\sqrt{6}} & 0 & 0 & 0 & 0 & 0 & 0 & 0 \\
0 & 0 & 0 & 0 & \frac{1}{\sqrt{6}} & 0 & 0 & 0 & 0 & -\frac{1}{\sqrt{2}} \\
0 & 0 & 0 & 0 & 0 & \frac{1}{2 \sqrt{3}} & 0 & 0 & 0 & 0 \\
0 & 0 & 0 & 0 & 0 & 0 & 0 & 0 & -\frac{1}{2} & 0
\end{array}
\right)
\end{eqnarray}

\begin{eqnarray}
T_5 = \left(
\begin{array}{llllllllll}
0 & 0 & 0 & 0 & 0 & -\frac{i}{2 \sqrt{3}} & 0 & 0 & 0 & 0 \\
0 & 0 & 0 & 0 & 0 & 0 & -\frac{i}{\sqrt{6}} & 0 & 0 & 0 \\
-\frac{i}{\sqrt{2}} & 0 & 0 & 0 & 0 & 0 & 0 & -\frac{i}{\sqrt{6}} &
0
& 0 \\
0 & -\frac{i}{\sqrt{3}} & 0 & 0 & 0 & 0 & 0 & 0 & -\frac{i}{2 \sqrt{3}} & 0 \\
0 & 0 & -\frac{i}{\sqrt{6}} & 0 & 0 & 0 & 0 & 0 & 0 & 0 \\
0 & 0 & 0 & 0 & \frac{i}{\sqrt{6}} & 0 & 0 & 0 & 0 &
\frac{i}{\sqrt{2}} \\
0 & 0 & 0 & 0 & 0 &  \frac{i}{2 \sqrt{3}} & 0 & 0 & 0 & 0 \\
0 & 0 & 0 & 0 & 0 & 0 & 0 & 0 & \frac{i}{2} & 0
\end{array}
\right)
\end{eqnarray}

\begin{eqnarray}
T_6 = \left(
\begin{array}{llllllllll}
0 & 0 & 0 & 0 & -\frac{1}{\sqrt{6}} & 0 & 0 & 0 & 0 & 0 \\
0 & 0 & 0 & 0 & 0 & -\frac{1}{2 \sqrt{3}} & 0 & 0 & 0 & 0 \\
0 & -\frac{1}{\sqrt{6}} & 0 & 0 & 0 & 0 & 0 & 0 & 0 & 0 \\
0 & 0 & -\frac{1}{\sqrt{3}} & 0 & 0 & 0 & 0 & \frac{1}{2 \sqrt{3}} &
0 & 0 \\
0 & 0 & 0 & -\frac{1}{\sqrt{2}} & 0 & 0 & 0 & 0 & \frac{1}{\sqrt{6}}
& 0 \\
0 & 0 & 0 & 0 & 0 & \frac{1}{2 \sqrt{3}} & 0 & 0 & 0 & 0 \\
0 & 0 & 0 & 0 & 0 & 0 & \frac{1}{\sqrt{6}} & 0 & 0 &
-\frac{1}{\sqrt{2}} \\
0 & 0 & 0 & 0 & 0 & 0 & 0 & \frac{1}{2} & 0 & 0
\end{array}
\right)
\end{eqnarray}

\begin{eqnarray}
T_7 = \left(
\begin{array}{llllllllll}
0 & 0 & 0 & 0 & \frac{i}{\sqrt{6}} & 0 & 0 & 0 & 0 & 0 \\
0 & 0 & 0 & 0 & 0 & \frac{i}{2 \sqrt{3}} & 0 & 0 & 0 & 0 \\
0 & -\frac{i}{\sqrt{6}} & 0 & 0 & 0 & 0 & 0 & 0 & 0 & 0 \\
0 & 0 & -\frac{i}{\sqrt{3}} & 0 & 0 & 0 & 0 & -\frac{i}{2 \sqrt{3}}
&
0 & 0 \\
0 & 0 & 0 & -\frac{i}{\sqrt{2}} & 0 & 0 & 0 & 0 &
-\frac{i}{\sqrt{6}}
& 0 \\
0 & 0 & 0 & 0 & 0 & \frac{i}{2 \sqrt{3}} & 0 & 0 & 0 & 0 \\
0 & 0 & 0 & 0 & 0 & 0 & \frac{i}{\sqrt{6}} & 0 & 0 &
\frac{i}{\sqrt{2}} \\
0 & 0 & 0 & 0 & 0 & 0 & 0 & -\frac{i}{2} & 0 & 0
\end{array}
\right)
\end{eqnarray}

\begin{eqnarray}
T_8 = \left(
\begin{array}{llllllllll}
0 & 0 & 0 & 0 & 0 & 0 & 0 & 0 & 0 & 0 \\
0 & 0 & 0 & 0 & 0 & 0 & 0 & 0 & 0 & 0 \\
0 & 0 & 0 & 0 & \frac{1}{\sqrt{2}} & 0 & 0 & 0 & 0 & 0 \\
0 & 0 & 0 & 0 & 0 & \frac{1}{\sqrt{2}} & 0 & 0 & 0 & 0 \\
0 & 0 & 0 & 0 & 0 & 0 & \frac{1}{\sqrt{2}} & 0 & 0 & 0 \\
0 & 0 & 0 & 0 & 0 & 0 & 0 & -\frac{1}{\sqrt{2}} & 0 & 0 \\
0 & 0 & 0 & 0 & 0 & 0 & 0 & 0 & -\frac{1}{\sqrt{2}} & 0 \\
0 & 0 & 0 & 0 & 0 & 0 & 0 & 0 & 0 & 0
\end{array}
\right)
\end{eqnarray}

\subsection{Decimet Generator 10x10 Matrices ${\bf F}_{(10)}^{a}$}
\label{sect:Decimet Generators}

\begin{eqnarray}
{\bf F}_{(10)}^{1} &=& \left(
\begin{array}{llllllllll|l}
 0 & \frac{\sqrt{3}}{2} & 0 & 0 & 0 & 0 & 0 & 0 & 0 & 0 & \Delta^{++} \\
 \frac{\sqrt{3}}{2} & 0 & 1 & 0 & 0 & 0 & 0 & 0 & 0 & 0 & \Delta^+ \\
 0 & 1 & 0 & \frac{\sqrt{3}}{2} & 0 & 0 & 0 & 0 & 0 & 0 & \Delta^0 \\
 0 & 0 & \frac{\sqrt{3}}{2} & 0 & 0 & 0 & 0 & 0 & 0 & 0 & \Delta^- \\
 0 & 0 & 0 & 0 & 0 & \frac{1}{\sqrt{2}} & 0 & 0 & 0 & 0 & \Sigma^{*+} \\
 0 & 0 & 0 & 0 & \frac{1}{\sqrt{2}} & 0 & \frac{1}{\sqrt{2}} & 0 & 0 & 0 & \Sigma^{*0} \\
 0 & 0 & 0 & 0 & 0 & \frac{1}{\sqrt{2}} & 0 & 0 & 0 & 0 & \Sigma^{*-} \\
 0 & 0 & 0 & 0 & 0 & 0 & 0 & 0 & \frac{1}{2} & 0 & \Xi^{*0} \\
 0 & 0 & 0 & 0 & 0 & 0 & 0 & \frac{1}{2} & 0 & 0 & \Xi^{*-} \\
 0 & 0 & 0 & 0 & 0 & 0 & 0 & 0 & 0 & 0 & \Omega \\
\hline \\
\Delta^{++} & \Delta^+ & \Delta^0 & \Delta^- &
\Sigma^{*+} & \Sigma^{*0} & \Sigma^{*-} & \Xi^{*0} & \Xi^{*-} &
\Omega
\end{array}
\right)
\end{eqnarray}

\begin{eqnarray}
{\bf F}_{(10)}^{2} &=& \left(
\begin{array}{llllllllll}
 0 & -\frac{i \sqrt{3}}{2} & 0 & 0 & 0 & 0 & 0 & 0 & 0 & 0
   \\
 \frac{i \sqrt{3}}{2} & 0 & -i & 0 & 0 & 0 & 0 & 0 & 0 & 0
   \\
 0 & i & 0 & -\frac{i \sqrt{3}}{2} & 0 & 0 & 0 & 0 & 0 & 0
   \\
 0 & 0 & \frac{i \sqrt{3}}{2} & 0 & 0 & 0 & 0 & 0 & 0 & 0 \\
 0 & 0 & 0 & 0 & 0 & -\frac{i}{\sqrt{2}} & 0 & 0 & 0 & 0 \\
 0 & 0 & 0 & 0 & \frac{i}{\sqrt{2}} & 0 &
   -\frac{i}{\sqrt{2}} & 0 & 0 & 0 \\
 0 & 0 & 0 & 0 & 0 & \frac{i}{\sqrt{2}} & 0 & 0 & 0 & 0 \\
 0 & 0 & 0 & 0 & 0 & 0 & 0 & 0 & -\frac{i}{2} & 0 \\
 0 & 0 & 0 & 0 & 0 & 0 & 0 & \frac{i}{2} & 0 & 0 \\
 0 & 0 & 0 & 0 & 0 & 0 & 0 & 0 & 0 & 0
\end{array}
\right)
\end{eqnarray}

\begin{eqnarray}
{\bf F}_{(10)}^{3} &=&\left(
\begin{array}{llllllllll}
 \frac{3}{2} & 0 & 0 & 0 & 0 & 0 & 0 & 0 & 0 & 0 \\
 0 & \frac{1}{2} & 0 & 0 & 0 & 0 & 0 & 0 & 0 & 0 \\
 0 & 0 & -\frac{1}{2} & 0 & 0 & 0 & 0 & 0 & 0 & 0 \\
 0 & 0 & 0 & -\frac{3}{2} & 0 & 0 & 0 & 0 & 0 & 0 \\
 0 & 0 & 0 & 0 & 1 & 0 & 0 & 0 & 0 & 0 \\
 0 & 0 & 0 & 0 & 0 & 0 & 0 & 0 & 0 & 0 \\
 0 & 0 & 0 & 0 & 0 & 0 & -1 & 0 & 0 & 0 \\
 0 & 0 & 0 & 0 & 0 & 0 & 0 & \frac{1}{2} & 0 & 0 \\
 0 & 0 & 0 & 0 & 0 & 0 & 0 & 0 & -\frac{1}{2} & 0 \\
 0 & 0 & 0 & 0 & 0 & 0 & 0 & 0 & 0 & 0
\end{array}
\right)
\end{eqnarray}

\begin{eqnarray}
{\bf F}_{(10)}^{4} &=& \left(
\begin{array}{llllllllll}
 0 & 0 & 0 & 0 & \frac{\sqrt{3}}{2} & 0 & 0 & 0 & 0 & 0 \\
 0 & 0 & 0 & 0 & 0 & \frac{1}{\sqrt{2}} & 0 & 0 & 0 & 0 \\
 0 & 0 & 0 & 0 & 0 & 0 & \frac{1}{2} & 0 & 0 & 0 \\
 0 & 0 & 0 & 0 & 0 & 0 & 0 & 0 & 0 & 0 \\
 \frac{\sqrt{3}}{2} & 0 & 0 & 0 & 0 & 0 & 0 & 1 & 0 & 0 \\
 0 & \frac{1}{\sqrt{2}} & 0 & 0 & 0 & 0 & 0 & 0 &
   \frac{1}{\sqrt{2}} & 0 \\
 0 & 0 & \frac{1}{2} & 0 & 0 & 0 & 0 & 0 & 0 & 0 \\
 0 & 0 & 0 & 0 & 1 & 0 & 0 & 0 & 0 & \frac{\sqrt{3}}{2} \\
 0 & 0 & 0 & 0 & 0 & \frac{1}{\sqrt{2}} & 0 & 0 & 0 & 0 \\
 0 & 0 & 0 & 0 & 0 & 0 & 0 & \frac{\sqrt{3}}{2} & 0 & 0
\end{array}
\right)
\end{eqnarray}

\begin{eqnarray}
{\bf F}_{(10)}^{5} &=& \left(
\begin{array}{llllllllll}
 0 & 0 & 0 & 0 & -\frac{i \sqrt{3}}{2} & 0 & 0 & 0 & 0 & 0
   \\
 0 & 0 & 0 & 0 & 0 & -\frac{i}{\sqrt{2}} & 0 & 0 & 0 & 0 \\
 0 & 0 & 0 & 0 & 0 & 0 & -\frac{i}{2} & 0 & 0 & 0 \\
 0 & 0 & 0 & 0 & 0 & 0 & 0 & 0 & 0 & 0 \\
 \frac{i \sqrt{3}}{2} & 0 & 0 & 0 & 0 & 0 & 0 & -i & 0 & 0
   \\
 0 & \frac{i}{\sqrt{2}} & 0 & 0 & 0 & 0 & 0 & 0 &
   -\frac{i}{\sqrt{2}} & 0 \\
 0 & 0 & \frac{i}{2} & 0 & 0 & 0 & 0 & 0 & 0 & 0 \\
 0 & 0 & 0 & 0 & i & 0 & 0 & 0 & 0 & -\frac{i \sqrt{3}}{2}
   \\
 0 & 0 & 0 & 0 & 0 & \frac{i}{\sqrt{2}} & 0 & 0 & 0 & 0 \\
 0 & 0 & 0 & 0 & 0 & 0 & 0 & \frac{i \sqrt{3}}{2} & 0 & 0
\end{array}
\right)
\end{eqnarray}

\begin{eqnarray}
{\bf F}_{(10)}^{6} &=&\left(
\begin{array}{llllllllll}
 0 & 0 & 0 & 0 & 0 & 0 & 0 & 0 & 0 & 0 \\
 0 & 0 & 0 & 0 & \frac{1}{2} & 0 & 0 & 0 & 0 & 0 \\
 0 & 0 & 0 & 0 & 0 & \frac{1}{\sqrt{2}} & 0 & 0 & 0 & 0 \\
 0 & 0 & 0 & 0 & 0 & 0 & \frac{\sqrt{3}}{2} & 0 & 0 & 0 \\
 0 & \frac{1}{2} & 0 & 0 & 0 & 0 & 0 & 0 & 0 & 0 \\
 0 & 0 & \frac{1}{\sqrt{2}} & 0 & 0 & 0 & 0 &
   \frac{1}{\sqrt{2}} & 0 & 0 \\
 0 & 0 & 0 & \frac{\sqrt{3}}{2} & 0 & 0 & 0 & 0 & 1 & 0 \\
 0 & 0 & 0 & 0 & 0 & \frac{1}{\sqrt{2}} & 0 & 0 & 0 & 0 \\
 0 & 0 & 0 & 0 & 0 & 0 & 1 & 0 & 0 & \frac{\sqrt{3}}{2} \\
 0 & 0 & 0 & 0 & 0 & 0 & 0 & 0 & \frac{\sqrt{3}}{2} & 0
\end{array}
\right)
\end{eqnarray}

\begin{eqnarray}
{\bf F}_{(10)}^{7} &=& \left(
\begin{array}{llllllllll}
 0 & 0 & 0 & 0 & 0 & 0 & 0 & 0 & 0 & 0 \\
 0 & 0 & 0 & 0 & -\frac{i}{2} & 0 & 0 & 0 & 0 & 0 \\
 0 & 0 & 0 & 0 & 0 & -\frac{i}{\sqrt{2}} & 0 & 0 & 0 & 0 \\
 0 & 0 & 0 & 0 & 0 & 0 & -\frac{i \sqrt{3}}{2} & 0 & 0 & 0 \\
 0 & \frac{i}{2} & 0 & 0 & 0 & 0 & 0 & 0 & 0 & 0 \\
 0 & 0 & \frac{i}{\sqrt{2}} & 0 & 0 & 0 & 0 & -\frac{i}{\sqrt{2}} & 0 & 0 \\
 0 & 0 & 0 & \frac{i \sqrt{3}}{2} & 0 & 0 & 0 & 0 & -i & 0 \\
 0 & 0 & 0 & 0 & 0 & \frac{i}{\sqrt{2}} & 0 & 0 & 0 & 0 \\
 0 & 0 & 0 & 0 & 0 & 0 & i & 0 & 0 & -\frac{i \sqrt{3}}{2} \\
 0 & 0 & 0 & 0 & 0 & 0 & 0 & 0 & \frac{i \sqrt{3}}{2} & 0
\end{array}
\right)
\end{eqnarray}

\begin{eqnarray}
{\bf F}_{(10)}^{8} &=& \left(
\begin{array}{llllllllll}
 \frac{\sqrt{3}}{2} & 0 & 0 & 0 & 0 & 0 & 0 & 0 & 0 & 0 \\
 0 & \frac{\sqrt{3}}{2} & 0 & 0 & 0 & 0 & 0 & 0 & 0 & 0 \\
 0 & 0 & \frac{\sqrt{3}}{2} & 0 & 0 & 0 & 0 & 0 & 0 & 0 \\
 0 & 0 & 0 & \frac{\sqrt{3}}{2} & 0 & 0 & 0 & 0 & 0 & 0 \\
 0 & 0 & 0 & 0 & 0 & 0 & 0 & 0 & 0 & 0 \\
 0 & 0 & 0 & 0 & 0 & 0 & 0 & 0 & 0 & 0 \\
 0 & 0 & 0 & 0 & 0 & 0 & 0 & 0 & 0 & 0 \\
 0 & 0 & 0 & 0 & 0 & 0 & 0 & -\frac{\sqrt{3}}{2} & 0 & 0 \\
 0 & 0 & 0 & 0 & 0 & 0 & 0 & 0 & -\frac{\sqrt{3}}{2} & 0 \\
 0 & 0 & 0 & 0 & 0 & 0 & 0 & 0 & 0 & -\sqrt{3}
\end{array}
\right)
\end{eqnarray}

\subsection{Singlet-Octet 1x8 Transition Matrices $T^a_{1/8}$}
\label{s:Veljko8x1}

\begin{eqnarray}
{\rm \bf T}^1_{1/8} = \left(
\begin{array}{llllllll|l}
0 & 0 & -{1\over\sqrt2} & 0 & {1\over\sqrt2} & 0 & 0 & 0 & \Lambda_1 \\
\hline \\
p & n & \Sigma^{+} & \Sigma^{0} & \Sigma^{-} & \Xi^{0} & \Xi^{-} &
\Lambda_8
\end{array}
\right)
\end{eqnarray}

\begin{eqnarray}
{\rm \bf T}^2_{1/8} = \left(
\begin{array}{llllllll}
0 & 0 & -{i\over\sqrt2} & 0 & -{i\over\sqrt2} & 0 & 0 & 0
\end{array}
\right)
\end{eqnarray}

\begin{eqnarray}
{\rm \bf T}^3_{1/8} = \left(
\begin{array}{llllllll}
0 & 0 & 0 & 1 & 0 & 0 & 0 & 0
\end{array}
\right)
\end{eqnarray}

\begin{eqnarray}
{\rm \bf T}^4_{1/8} = \left(
\begin{array}{llllllll}
{1\over\sqrt2} & 0 & 0 & 0 & 0 & 0 & -{1\over\sqrt2} & 0
\end{array}
\right)
\end{eqnarray}

\begin{eqnarray}
{\rm \bf T}^5_{1/8} = \left(
\begin{array}{llllllll}
{i\over\sqrt2} & 0 & 0 & 0 & 0 & 0 & {i\over\sqrt2} & 0
\end{array}
\right)
\end{eqnarray}

\begin{eqnarray}
{\rm \bf T}^6_{1/8} = \left(
\begin{array}{llllllll}
0 & {1\over\sqrt2} & 0 & 0 & 0 & {1\over\sqrt2} & 0 & 0
\end{array}
\right)
\end{eqnarray}

\begin{eqnarray}
{\rm \bf T}^7_{1/8} = \left(
\begin{array}{llllllll}
0 & {i\over\sqrt2} & 0 & 0 & 0 & -{i\over\sqrt2} & 0 & 0
\end{array}
\right)
\end{eqnarray}

\begin{eqnarray}
{\rm \bf T}^8_{1/8} = \left(
\begin{array}{llllllll}
0 & 0 & 0 & 0 & 0 & 0 & 0 & 1
\end{array}
\right)
\end{eqnarray}

\end{document}